\documentclass[aps,prb,reprint]{revtex4-2}
\usepackage{graphicx} 
\usepackage{amsmath} 
\usepackage{amsfonts}
\usepackage{bm} 
\usepackage{natbib}
\usepackage{natmove}
\usepackage[colorlinks=true, linkcolor=blue, citecolor=blue, urlcolor=blue, linktoc=page, bookmarks=false, pdfstartview={FitH}, pdfborder={0 0 0.0 [3 3]}]{hyperref} 
\usepackage{color} 
\usepackage{dcolumn} 
\newcolumntype{.}{D{.}{.}{0.3}}
\newcolumntype{-}{D{.}{.}{4.0}}
\usepackage{makecell}
\usepackage{multirow}
\usepackage{cleveref}
\usepackage{easyReview} 
\crefname{figure}{Fig.}{Figs}
\crefname{table}{Table}{Tables}
\crefname{section}{Sec.}{Secs.}
\crefname{equation}{Eq.}{Eqs.}
\usepackage[mathscr]{euscript}
\newcommand\identity{1\kern-0.25em\text{l}}
\usepackage{braket}
\renewcommand{\today}{\number\day \space \ifcase \month \or January\or February\or March\or April\or May\or June\or July\or August\or September\or October\or November\or December\fi \space \number\year} 
\def\m1r{\multicolumn{1}{r}}
\begin{document}
\title{Spin-orbit interaction, band topology, and spin texture in BiInO$_\text{3}$(001) surface}
\author{Ramsamoj \surname{Kewat}}
\email{ramsamoj18@iiserb.ac.in}
\author{Nirmal \surname{Ganguli}}
\email{NGanguli@iiserb.ac.in}
\affiliation{Department of Physics, Indian Institute of Science Education and Research Bhopal, Bhauri, Bhopal 462066, India}
\date{\today}
\begin{abstract}
This research investigates the implication of spin-orbit interaction (SOI) and symmetry on the band topology and spin texture at the (001) surface of BiInO$_3$. Using density functional theory (DFT) and symmetry analysis, the study explores the impact of surface termination on the electronic structure, particularly focusing on how the loss of translational symmetry at the surface influences the band topology of the surface states. Key findings include discovering two dangling surface states (named SS1 and SS2) with distinct spin textures. SS1 exhibits Rashba spin splitting due to surface inversion asymmetry, characterised by isotropic effective mass and tangential spin alignment on constant energy contours. In contrast, SS2 features a persistent spin Texture (PST) spin texture, a momentum-independent spin polarisation. The orbital contributions to these bands, dominated by specific s and p orbitals, dictate the direction and nature of the spin texture. This study highlights BIO's potential as a platform for spintronic applications, where control over spin textures and electronic properties at the surface can enable the design of advanced spin-based devices. The findings bridge the gap between symmetry-driven theoretical frameworks and practical material functionalities, offering insights into the interplay of surface symmetry, SOI, and band topology.
\end{abstract}
\maketitle

\section{\label{sec:intro}Introduction}

\begin{figure*}[t] 
    \graphicspath{ {/}}
    \includegraphics[scale=0.26]{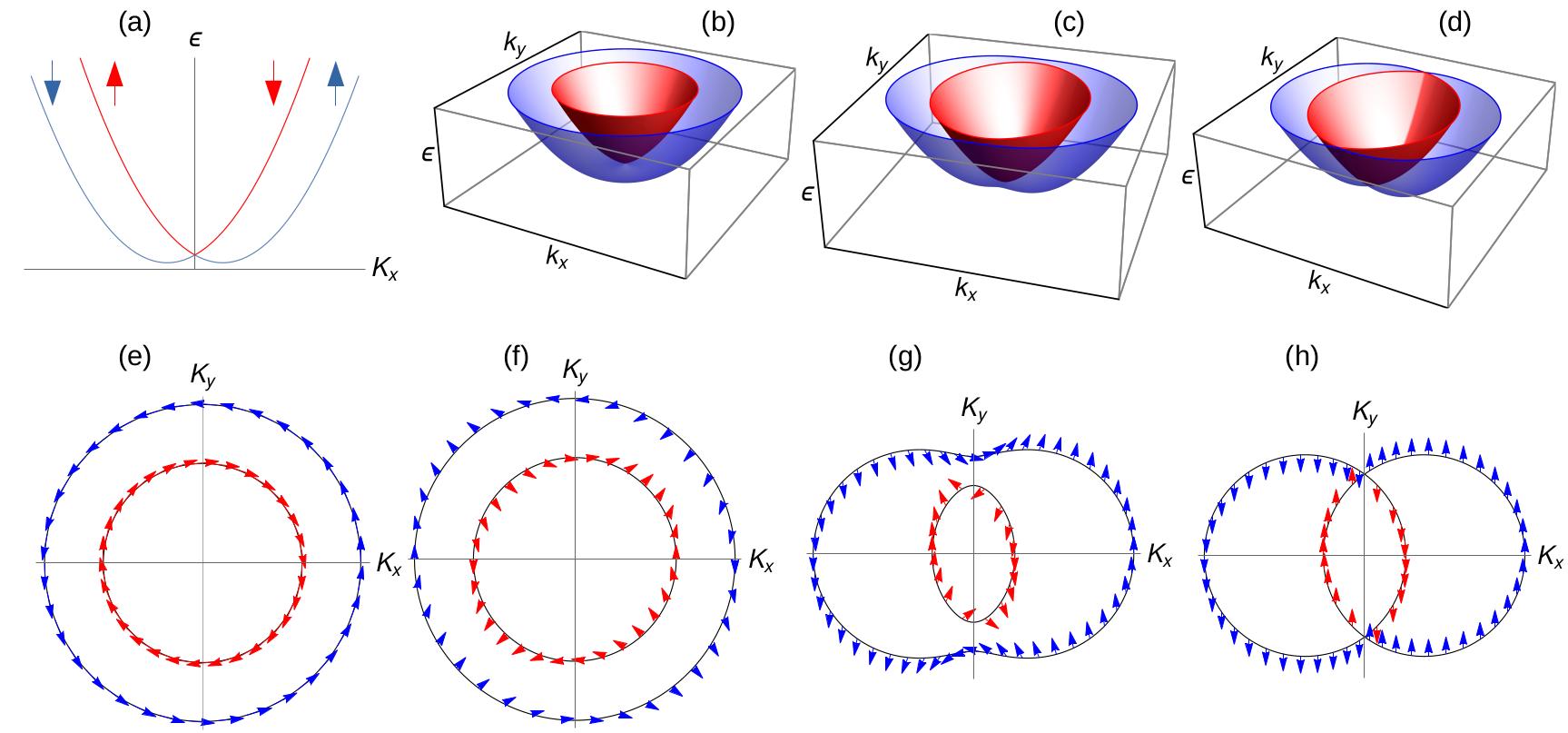}
    \caption{\label{fig:RD_image}
(a) Two-dimensional Rashba or Dresselhaus spin–orbit interaction (SOI) band splitting, where arrows indicate the spin direction.  
(b)–(d) Three-dimensional band structures illustrating: (b) either Rashba or Dresselhaus SOI, (c) combined Rashba and Dresselhaus SOI with unequal strengths ($\alpha_R \neq \alpha_D \neq 0$), and (d) Persistent Spin Texture (PST) bands where $\alpha_R = \alpha_D$.  
(e)–(h) Spin textures projected on constant energy contours: (e) Rashba spin texture, (f) Dresselhaus spin texture, (g) mixed Rashba and Dresselhaus texture with $\alpha_R \neq \alpha_D \neq 0$, and (h) PST spin texture.
}

\end{figure*}

To explore and understand the electronic properties of materials, it is essential to consider their topological nature\cite{Kane2005, Moore2007, Fu2006, Chiu2016, Fu2007}. The behavior of electrons at surfaces and boundaries can differ significantly from that in the bulk due to various symmetries, such as crystal symmetry and internal symmetries like time-reversal symmetry ($\mathcal{T}$), chiral symmetry ($\mathcal{S}$), and charge conjugation symmetry ($\mathcal{C}$). When van der Waals (vdW) forces or interatomic bonds between layers are broken, it can lead to trivial surface states caused by dangling bonds\cite{achal2018lithography,berthe2008probing,boland1993manipulating}. These so-called dangling bands are highly reactive and can trap charge carriers (electrons or holes), thereby affecting the transport properties at the material’s surface or interface\cite{lin2013topological}.
Moreover, the way energy bands disperse in a material is also shaped by its crystal symmetry. The interplay between time-reversal symmetry and crystal symmetry determines whether certain bands remain degenerate or split in time-reversal symmetry-preserving materials. While translational symmetry is common to all crystals, some crystals also possess mirror or rotational symmetry combined with partial translation; these are known as non-symmorphic crystals. A mirror combined with a partial translation leads to glide reflection symmetry, while rotation combined with non-primitive translation gives rise to screw rotational symmetry.
Materials with such non-symmorphic symmetries, such as KHgX (where X = As, Sb, Bi),\cite{Wang2016b,Ma2017} can host interesting band structures at the surface. These include hourglass-shaped band dispersions where four bands are connected in such a way that the Kramers partners exchange as one moves across the Brillouin zone. The intersection point of these hourglass bands is known as a Kramers–Weyl point \cite{Wang2016b, Ma2017, Wang2017, Wang2019, PhysRevB.96.075110, PhysRevB.94.195109, liu2023dirac}.
In certain momentum-space paths, space-time symmetries (such as time-reversal followed by glide mirror or screw rotation) enforce the degeneracy of Kramers partners along a line, resulting in a Kramers nodal line\cite{Xie2021}. These exotic fermions have been observed not only on surfaces~\cite{wieder2018wallpaper}, but also in the bulk and two-dimensional materials\cite{Wang2019}. When topological band crossings arise due to symmorphic crystal symmetry and internal symmetry, they lead to Dirac or Weyl points. However, these crossings are usually sensitive to external conditions like electric or magnetic fields, making them less stable for practical applications. In contrast, band crossings that emerge from non-symmorphic symmetries, such as those seen in hourglass, accordion\cite{zhang2019catalogue}, or M\"obius fermion states\cite{Zhang2020}, are much more robust. More complex symmetry operations protect these and do not break easily under small perturbations, making them more promising for real-world applications.
This robustness inspires the study of spin textures in such band structures. In non-symmorphic crystals, the absence of inversion symmetry gives rise to an effective momentum-dependent magnetic field $\vec{B}$ within the crystal. This field couples with the atomic angular momentum to produce spin–orbit interaction (SOI), which lifts the degeneracy of bands away from time-reversal invariant momenta (TRIM), even in the absence of external fields\cite{ishizaka2011giant,manchon2015new}.
The classic works of Dresselhaus and Rashba describe that this SOI leads to effective Hamiltonians $\mathcal{H}_{\text{so}} = \vec{B} \cdot \vec{\sigma}$, where $\vec{B}$ depends on the crystal symmetry. Depending on the point group and the little group \cite{bradley1972symmetry,dresselhaus2008group}at a high-symmetry point, $\vec{B}$, it can be linear, cubic, or even higher-order in $\vec{k}$~\cite{PhysRevLett.125.216405, PhysRevB.84.121401}. For gyrotropic crystals like those with point groups $\mathcal{D}_{2h}$ and $\mathcal{C}_{2v}$, the Rashba and Dresselhaus spin–orbit fields are linear in momentum: $\vec{B} = \alpha_R(k_x, -k_y)$ for Rashba~\cite{Bychkov1984}, and $\vec{B} = \alpha_D(k_x, k_y)$ for Dresselhaus\cite{1001954}.
The spin textures associated with these interactions are also momentum dependent. In Rashba systems, the spin is tangential to the constant energy contours, while in Dresselhaus systems, it is a mix of radial and tangential directions. The Rashba and Dresselhaus spin texture is shown in \cref{fig:RD_image}(e) and \cref{fig:RD_image}(f) respectively. This momentum dependence introduces complications in spintronics applications, especially for devices like spin-FETs, because spin relaxation can occur quickly due to scattering with impurities or defects, a process known as Dyakonov–Perel spin relaxation\cite{Dyakonov2017}.
To overcome these challenges, researchers look for materials exhibiting Persistent Spin Texture (PST) as shown in \cref{fig:RD_image}(h), a state where the spin orientation remains uniform and does not precess due to changes in momentum. PST can emerge when Rashba and Dresselhaus interactions contribute equally. This leads to a spin orientation that is robust against scattering and ideal for non-ballistic spintronic devices.

BiInO$_3$ is a particularly exciting material in this context. It is a non-symmorphic crystal that not only hosts hourglass fermions in the bulk but also features Kramers nodal lines\cite{xie2021kramers}. On its [001] surface, BiInO$_3$ shows dangling bands, and the corresponding band dispersion is governed by crystal symmetries present at the surface. The combination of mirror symmetry with a fractional in-plane translation gives rise to hourglass fermions even in these surface states. This makes BiInO$_3$ a promising candidate for hosting hourglass fermions with persistent spin textures in the dangling bands, offering great potential for spintronic and quantum device applications.


\section{\label{sec:method}Crystal Structure and Method}
\begin{figure*}
    \centering
    \graphicspath{ {/}}    
    \includegraphics[scale=0.5]{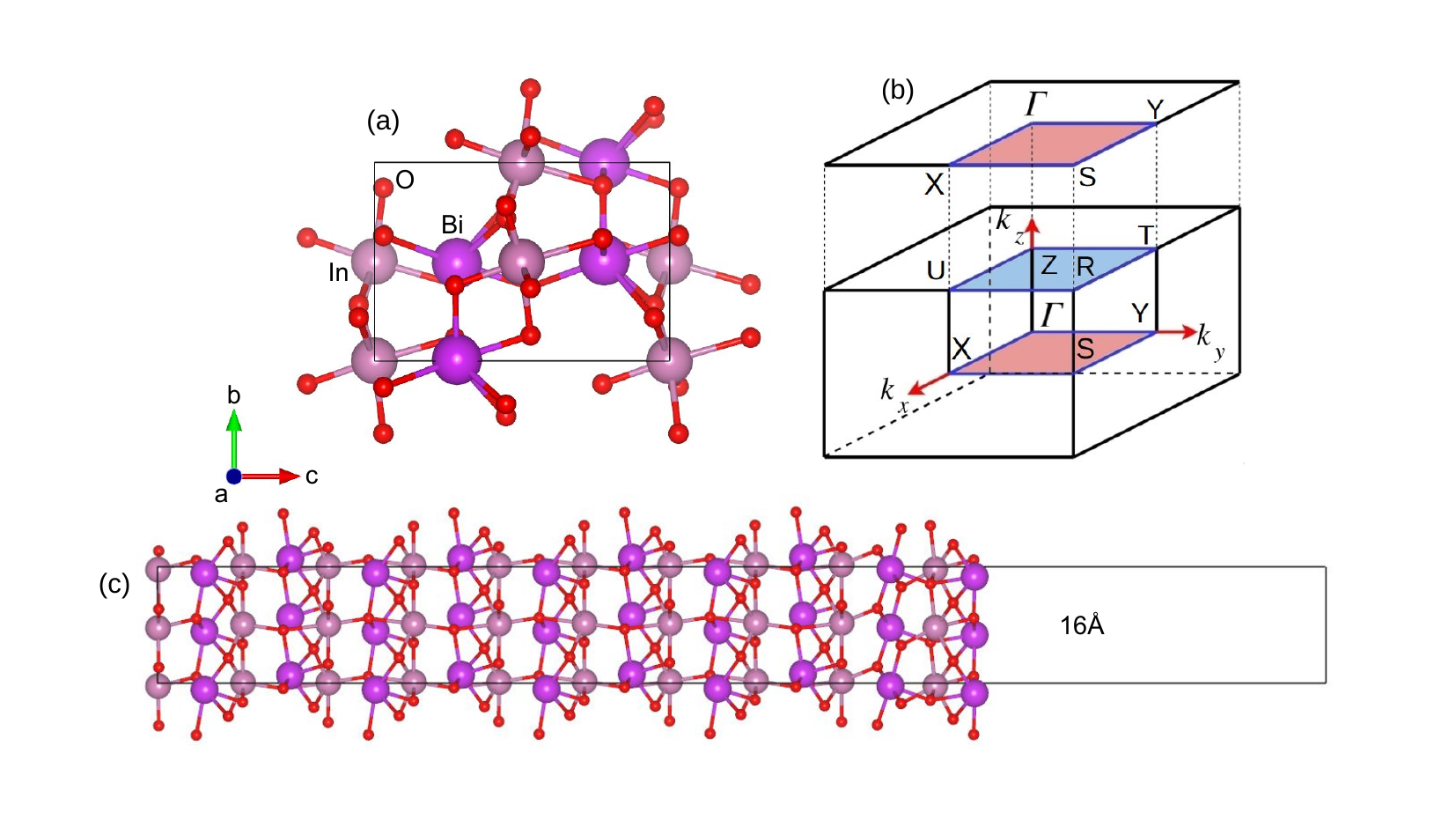}
 
    \caption{\label{fig:BiIn_surface}
(a) Crystal structure of bulk BiInO$_3$.  
(b) Surface Brillouin zone and high-symmetry path used in DFT calculations. The high-symmetry points for the surface of BiInO$_3$ are labeled as $\Gamma(0,0,0)$, $X(\pi,0,0)$, $S(\pi,\pi,0)$, and $Y(0,\pi,0)$.  
(c) A $1 \times 1 \times 5$ supercell of BiInO$_3$ with a vacuum layer of 16~\AA{} is constructed to model the [001] surface.
}
\end{figure*}

BiInO$_3$ crystallises in a non-centrosymmetric polar orthorhombic structure belonging to the space group $Pna2_1$ (No.~33) \cite{Belik2006}. This space group includes three key symmetry operations in addition to the identity \identity: a twofold screw rotation about the $z$-axis, denoted as $\Tilde{\mathcal{C}}_{2z}$ (\{$2_{001}|0~0~\frac{1}{2}$\} in Seitz notation); a glide reflection across the $x = 0$ plane, $\Tilde{\mathcal{M}}_x$ (\{$m_{100}|\frac{1}{2}~\frac{1}{2}~\frac{1}{2}$\}); and a glide reflection across the $y = 0$ plane, $\Tilde{\mathcal{M}}_y$ (\{$m_{010}|\frac{1}{2}~\frac{1}{2}~0$\}). It is important to note that the $\Tilde{\mathcal{C}}_{2z}$ and $\Tilde{\mathcal{M}}_x$ symmetries are broken when considering a BiInO$_3$ slab terminated along the (001) surface. The absence of inversion symmetry in this polar space group gives rise to a significant spontaneous electric polarisation along the $z$-axis, which is primarily driven by the displacement of heavy Bi ions and the structural distortion of BiO$_6$ and InO$_6$ octahedra.
The electronic structure calculations in this work are carried out using density functional theory (DFT) within the projector augmented wave (PAW) method \cite{paw}, employing a plane-wave basis set as implemented in the {\scshape VASP} package \cite{vasp1, vasp2}. A kinetic energy cutoff of 500~eV is used for the plane-wave expansion of wavefunctions. The exchange-correlation effects are treated using the generalised gradient approximation (GGA) in the form proposed by Perdew, Burke, and Ernzerhof (PBE) \cite{pbe}, for both bulk and slab calculations. For slab geometry, a five-unit-cell-thick BiInO$_3$ slab is constructed with a vacuum layer of 16~\AA\ along the $c$-axis to eliminate spurious interactions between periodic images, as illustrated in \cref{fig:BiIn_surface}(d). Brillouin zone integrations are performed using $\Gamma$-centred Monkhorst-Pack $k$-meshes: $11 \times 13 \times 9$ for the bulk and $11 \times 11 \times 1$ for the slab, evaluated using the corrected tetrahedron method \cite{BlochlPRB94T}. Both bulk and slab structures are relaxed until the Hellmann–Feynman forces on all atoms are less than $10^{-2}$~eV/\AA, and the stress on the lattice vectors is minimised.

 
\section{Results and discussions}
\subsection{Surface states in BiInO$_3$}
\begin{figure}[t!]
    \centering
    \includegraphics[width=0.9\linewidth]{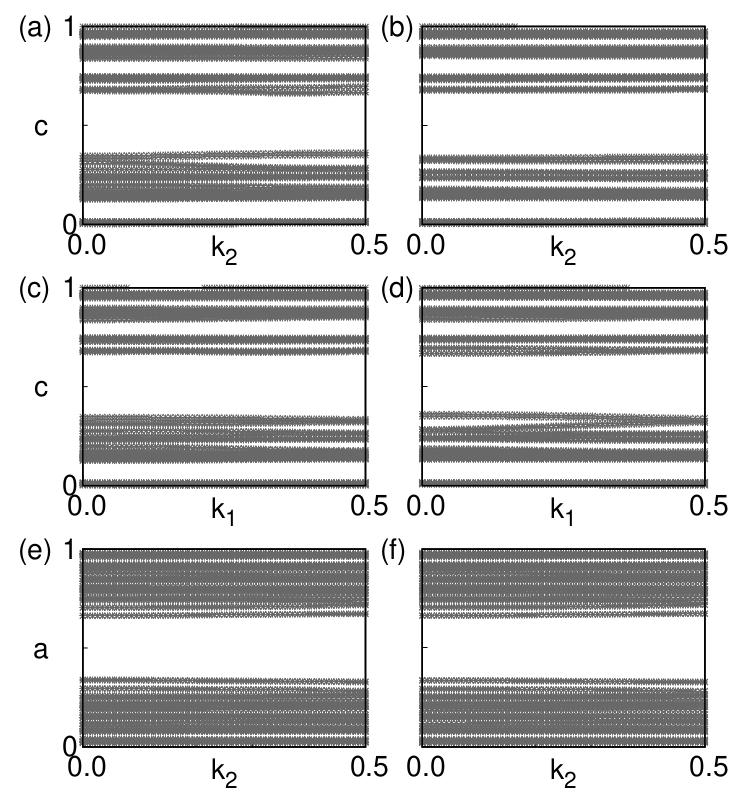}
    \caption{\label{fig:wc3D_z2.jpg}
Wilson loop (Wannier charge center) calculations on six time-reversal invariant momentum planes:  
(a) $k_1 = 0.0$, $z_2 = 0$;  
(b) $k_1 = 0.5$, $z_2 = 0$;  
(c) $k_2 = 0.0$, $z_2 = 0$;  
(d) $k_2 = 0.5$, $z_2 = 0$;  
(e) $k_3 = 0.0$, $z_2 = 0$;  
(f) $k_3 = 0.5$, $z_2 = 0$.  
The topological $\mathcal{Z}_2$ indices of BiInO$_3$ are given by $\nu = (0;000)$, indicating a trivial topological phase\cite{soluyanov2011computing}.}

\end{figure}

We have calculated the bulk band structure of BiInO$_3$, a semiconductor with a band gap of 2.77 eV without Spin-Orbit Interaction (SOI) and 2.55 eV with SOI. While calculating its surface band, we observed two pairs of surface bands. However, these bands do not connect the valence and conduction bulk bands, likely indicating trivial dangling bond\cite{lin2013topological}. To determine if these are trivial or non-trivial topological dangling bonds, we investigated the topological electronic properties of this material. Using WannierTools\cite{WU2017} and Wannier90\cite{MOSTOFI20142309}, we calculated the Wilson loop (Wannier charge centre) at six-time reversal invariant planes. The results showed topological $\mathcal{Z}_2$ indices, $\nu=(0;000)$ as shown in \cref{fig:wc3D_z2.jpg}, indicating that BiInO$_3$ is a trivial semiconductor according to the Fu and Kane classification scheme\cite{Fu2007}. Since $\mathcal{Z}_2$ for BiInO$_3$, no surface termination will have a topologically non-trivial surface band. This superstructure comprises five unit cells stacked along the c-direction, coupled by van der Waals (vdW) forces. The breaking of structural symmetry due to the vacuum created at the InO$_2$ and BiO terminating planes breaks the vdW coupling forces. This spoils the van der Waals (vdW) coupling force, which may lead to surface reconstruction and dangling bond formation. We study the detailed band dispersion of these dangling bonds from a symmetry perspective, focusing on how SOI and symmetry influence the band dispersion and spin texture of the bands in the following sections.

\subsection{Symmetries in BiInO$_3$}
\begin{figure*} [t!]
    \graphicspath{ {/}}
    \includegraphics[scale=0.65]{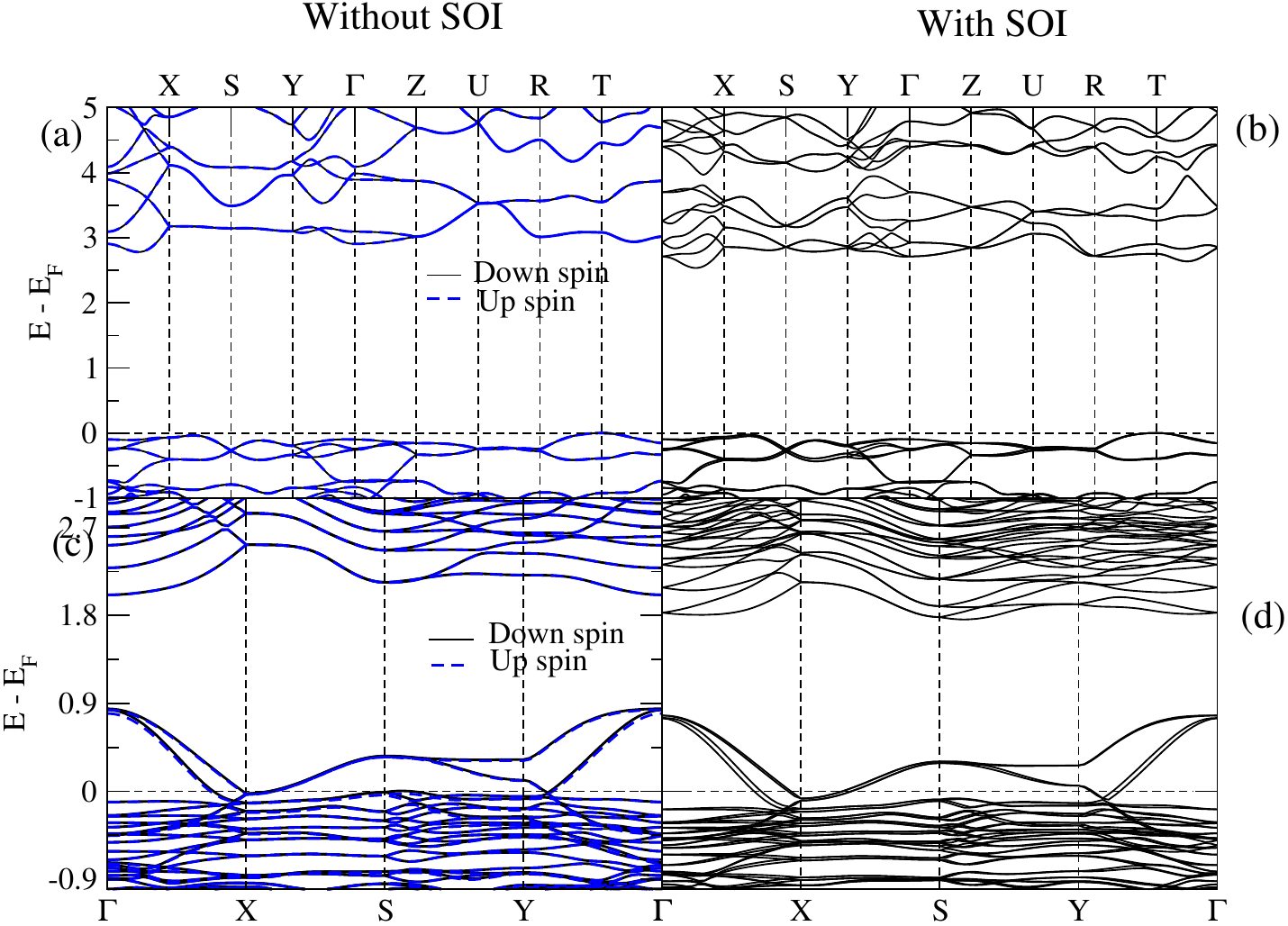}
    \caption{\label{fig:biband}(a) and (b) Band structure of Bulk BiInO$_3$ without SOI and with SOI, respectively. (c) and (d) Band structure of surface BiInO$_3$ without SOI and with SOI, respectively.}
\end{figure*}
\begin{figure}
    \centering
    \includegraphics[width=1\linewidth]{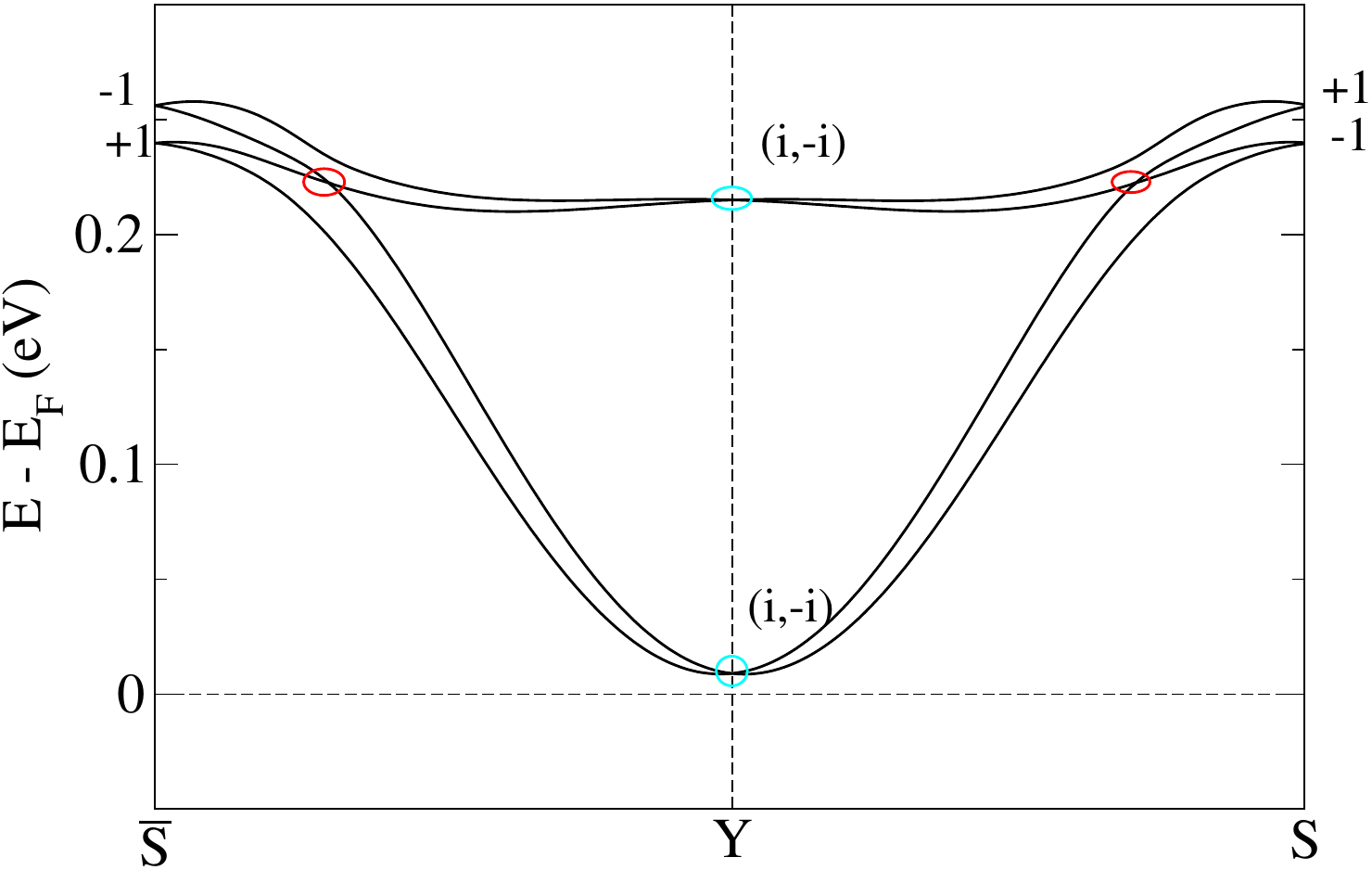}
    \caption{
Zoomed-in view of the hourglass band network in the surface state along the high-symmetry path $\bar{S} \rightarrow Y\rightarrow S$. The red-circled crossing point marks the hourglass band crossing, while the cyan-circled crossing corresponds to the Kramers–Weyl point.
}
  \label{fig:sysbandsmall}
\end{figure}
\begin{figure}
    \centering
    \includegraphics[width=1\linewidth]{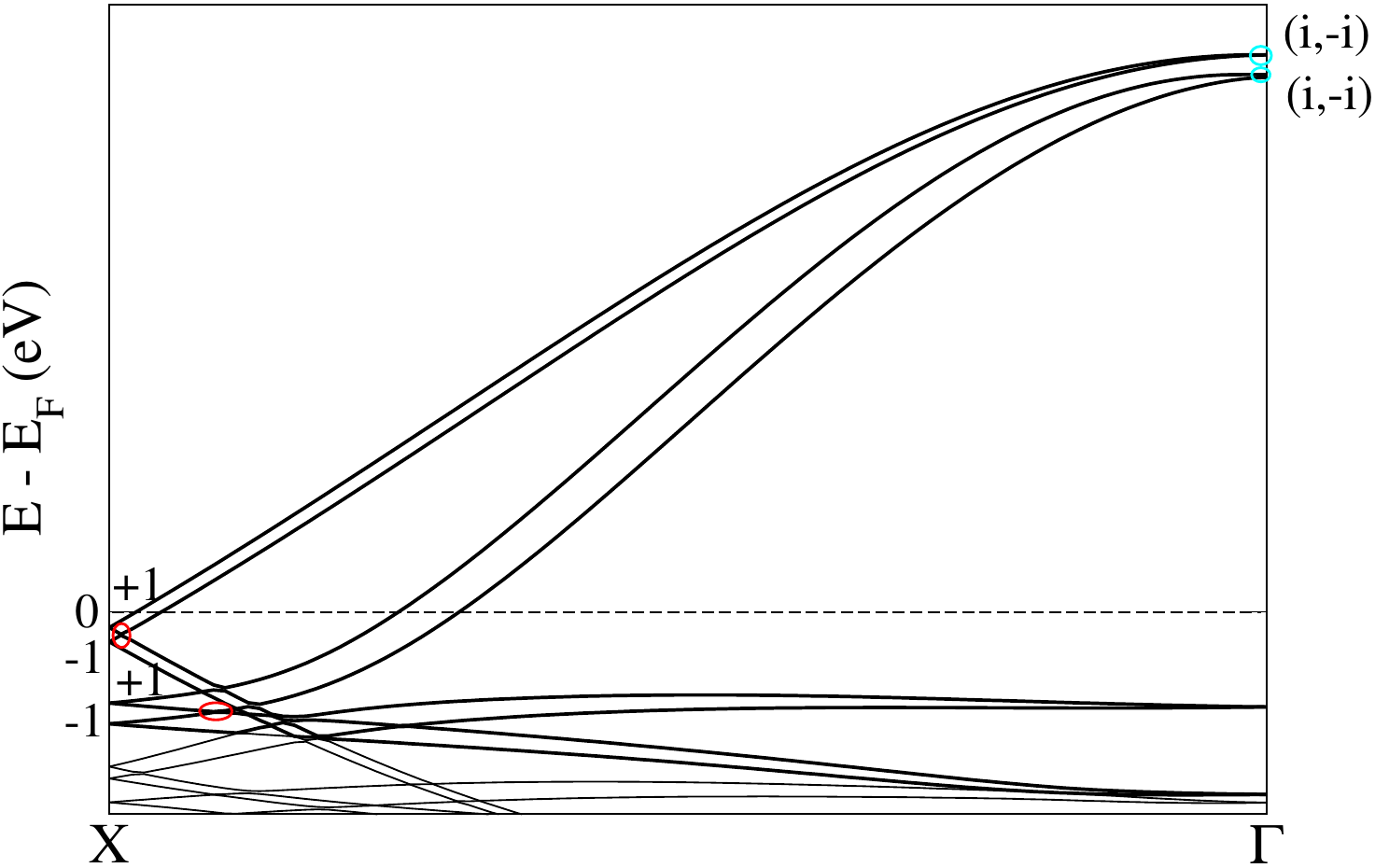}
    \caption{
Zoomed-in view of the hourglass band network in the surface state along the high-symmetry path $X \rightarrow \Gamma$. The red-circled crossing point indicates the hourglass band crossing, while the cyan-circled point corresponds to the Kramers–Weyl point.
}
\label{fig:xgxband}
\end{figure}

BiInO$_3$ is a non-symmorphic material with orthorhombic perovskite structure, with space group Pna2$_1$ (No.33)\cite{aroyo2006bilbao}. This material manifests various crystal symmetries, including the Identity symmetry (\identity) and time-reversal symmetry ($\mathcal{T}$), contributing to its unique topological characteristics. The following are the symmetries present in BiInO$_3$.

(1) Glide reflection Symmetry about yz-plane,

\begin{align} \label{eq:2}
\left\{ m_{100} \,\middle|\, \tfrac{1}{2}~\tfrac{1}{2}~\tfrac{1}{2} \right\}: (x, y, z) \rightarrow \left( -x + \tfrac{1}{2},\, y + \tfrac{1}{2},\, z + \tfrac{1}{2} \right)
\end{align}

(2) Glide reflection Symmetry about the xz-plane,
\begin{align} \label{eq:3}
\left\{ m_{010} \,\middle|\, \tfrac{1}{2}~0~\tfrac{1}{2} \right\}: (x, y, z) \rightarrow \left( x + \tfrac{1}{2},\, -y + \tfrac{1}{2},\, z \right)
\end{align}

(3) Two-fold screw rotation symmetry about the z-axis,
\begin{align} \label{eq:4}
\left\{ 2_{001} \,\middle|\, 0~0~\tfrac{1}{2} \right\}: (x, y, z) \rightarrow \left( -x,\, -y,\, z + \tfrac{1}{2} \right)
\end{align}

\textbf{ Surface Symmetry analysis:}
Since at 001 surfaces of BiInO$_3$, there is a restriction of translation along the (001)-direction, so,

both screw rotation symmetry $\left\{ 2_{001} \mid 0~0~\frac{1}{2} \right\}$ and Glide reflection about yz-plane $\left\{ m_{100} \mid \frac{1}{2}~\frac{1}{2}~\frac{1}{2} \right\}$
 spoilt, the only symmetry that is alive is Glide 
reflection symmetry about xz-plane $\left\{ m_{010} \mid \frac{1}{2}~0~\frac{1}{2} \right\}$ , having a double glide perpendicular to the z-axis. As the surface is created, no magnetic behaviour is going to change. Time reversal symmetry $\mathcal{T}$ still preserves, and identity symmetry $\identity$ exists, which is trivial. The space group's reduction results in the point group's lowering from `$\mathcal{C}_{2v}$' to `m', having xz-mirror symmetry $\mathcal{M}_y\left\{ m_{010} \mid 0~0~0 \right\}$ only. These symmetry changes at the surface hosting different  band structue at the surface than the bulk BiInO$_3$ 

The crystallographic symmetry inherent in materials significantly influences their electronic structure. In the case of BiInO$_3$, the interplay between non-symmorphic symmetries and Spin-Orbit Interaction (SOI) is pivotal in determining the band structure, giving rise to phenomena such as band degeneracy and splitting. Our focus is understanding how specific symmetries, such as glide reflection symmetry and time reversal, contribute to forming Kramers-like degeneracy along distinct high symmetry paths in momentum space, as illustrated in \cref{fig:biband}.

For our investigation, we've chosen a high symmetry path in momentum space, denoted as $\Gamma$$\rightarrow X$$\rightarrow S $$\rightarrow Y$$\rightarrow \Gamma$, on the surface for calculating energy bands. Before delving into the Eigenvalues and Eigenstates along these paths, it's necessary to comprehend how glide reflection and time reversal operators act on the Bloch Eigenstates.
 
BiInO$_3$ is non-magnetic, preserving time inversion symmetry and introducing spin-orbit interaction becomes a key aspect of our study. The time inversion operator, being antiunitary for Fermions, is a crucial element in understanding the band degeneracy and splitting along the chosen high symmetry path in the momentum space of this material. The time inversion operator is written as,

\begin{equation} \label{eq:5}
\mathcal{T}= \textit{i}\sigma_y\hat{\mathcal{K}}
\end{equation}
\begin{equation} \label{eq:6}
\mathcal{T}^2=\textit{i}^2\sigma_y^2\hat{\mathcal{K}}^2=-\identity
\end{equation}
where $\sigma_y$ Pauli spin matrix and $\hat{\mathcal{K}}$ is a complex conjugation operator in Bloch State.
The time inversion operator commutes with all the space group transformations and considers spin-orbit interaction and all the space group symmetry also present in spin space. Let us define space-time reversal symmetry, $\mathcal{T}\Tilde{\mathcal{M}_x}=\mathcal{T}\left\{ m_{100} \mid \frac{1}{2}~\frac{1}{2}~\frac{1}{2} \right\}$ and $\mathcal{T}\Tilde{\mathcal{M}_y}=\mathcal{T}\left\{ m_{010} \mid \frac{1}{2}~0~\frac{1}{2} \right\}$ also preserved in the crystal and particular section of momentum space keeping momentum path invariant. The Glide reflection symmetry, Time inversion symmetry, and their product can operate on Bloch states with crystal and spin space. Let us define $\psi(\vec{k})$ be the Bloch  state, and we can write the eigenvalue equation as;
\begin{equation} \label{eq:7}
\left\{ m_{100} \mid \frac{1}{2}~\frac{1}{2}~\frac{1}{2} \right\} \psi(\vec{k})^{\pm}=\pm\textit{i}e^{-\textit{i}(k_y+k_z)/2}\psi(\vec{k})^{\pm}
\end{equation}
\begin{equation}\label{eq:8}
\left\{ m_{010} \mid \frac{1}{2}~0~\frac{1}{2} \right\}\psi(\vec{k})^{\pm}=\pm\textit{i}e^{-\textit{i}k_x/2}\psi(\vec{k})^{\pm}
\end{equation}
\begin{equation}\label{eq:9}
\mathcal{T}\left\{ m_{100} \mid \frac{1}{2}~\frac{1}{2}~\frac{1}{2} \right\}\psi(\vec{k})^{\pm}=\pm e^{-\textit{i}(k_y+k_z)/2}\psi(\vec{k})^{\pm}
\end{equation}
\begin{equation}\label{eq:10}
\mathcal{T}\left\{ m_{010} \mid \frac{1}{2}~0~\frac{1}{2} \right\}\psi(\vec{k})^{\pm}=\pm e^{-\textit{i}k_x/2}\psi(\vec{k})^{\pm}
\end{equation}

The $\pm$ and $\pm i$ came from the spin space in the momentum-dependent eigenvalue in the above eigenvalue equation.

\begin{figure*}
    \graphicspath{ {/}}
    \includegraphics[scale=0.6]{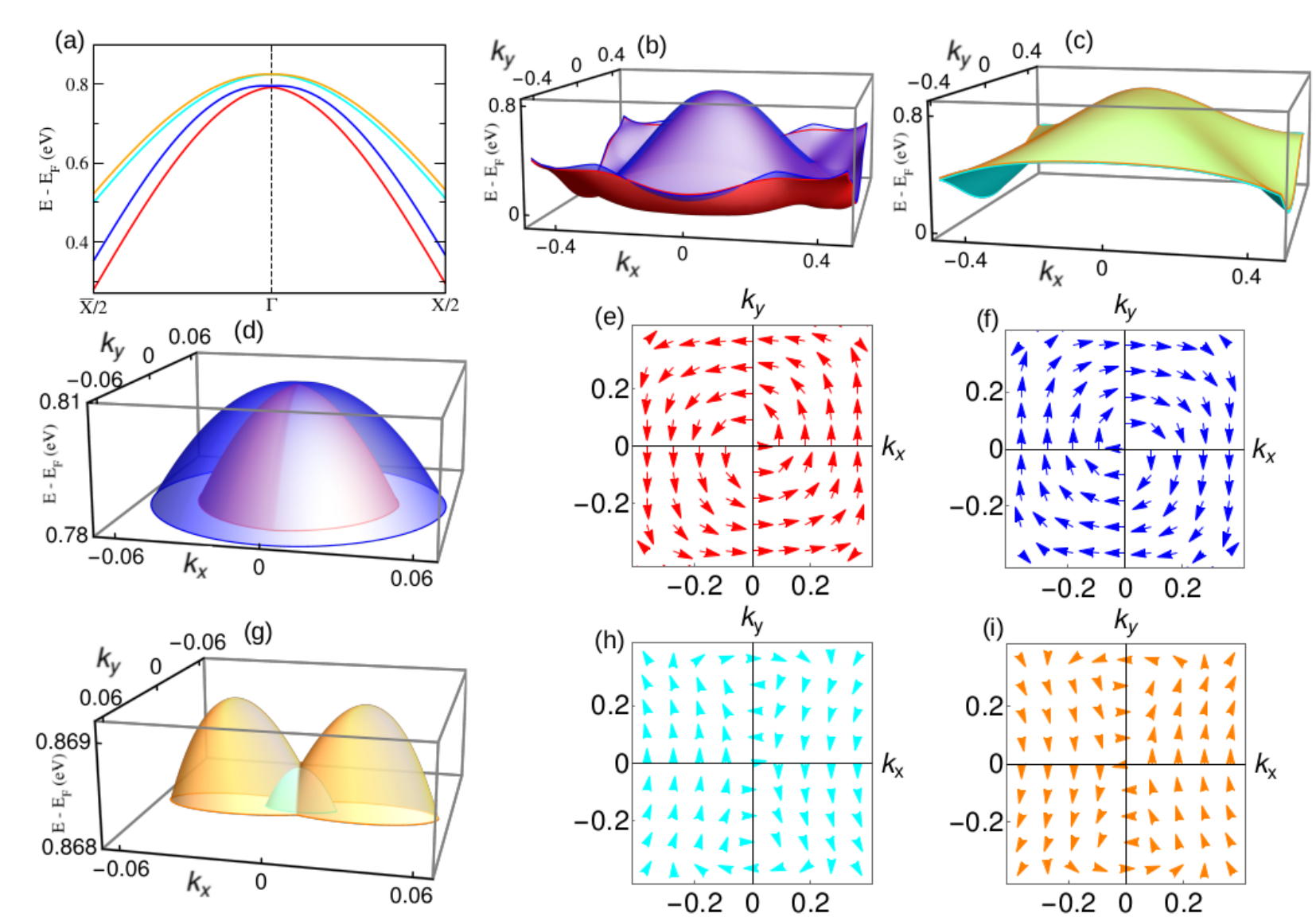}
    \caption{\label{fig:fermird}
(a) Energy band of surface electrons along the high-symmetry path $\bar{X}/2 \rightarrow \Gamma \rightarrow X/2$. The lower pair of bands shows Rashba spin–orbit interaction (SOI), while the upper pair has both Rashba and Dresselhaus SOI of equal strength, forming a persistent spin texture (PST) state. 
(b) and (c) show the 3D energy bands of the lower and upper pairs, respectively. 
(d) 3D Rashba band zoomed in around the $\Gamma$ point in a narrow energy range. 
(e) and (f) Spin textures of the inner and outer Rashba bands, respectively, calculated using density functional theory (DFT). 
(g) PSH band shown in a narrow energy window near the $\Gamma$ point. 
(h) and (i) Spin textures of the inner and outer PSH bands, respectively, also obtained from DFT calculations.}

\end{figure*}

 \begin{figure*}[!t]
    \graphicspath{ {/}}
    \centering
   \includegraphics[scale=0.5]{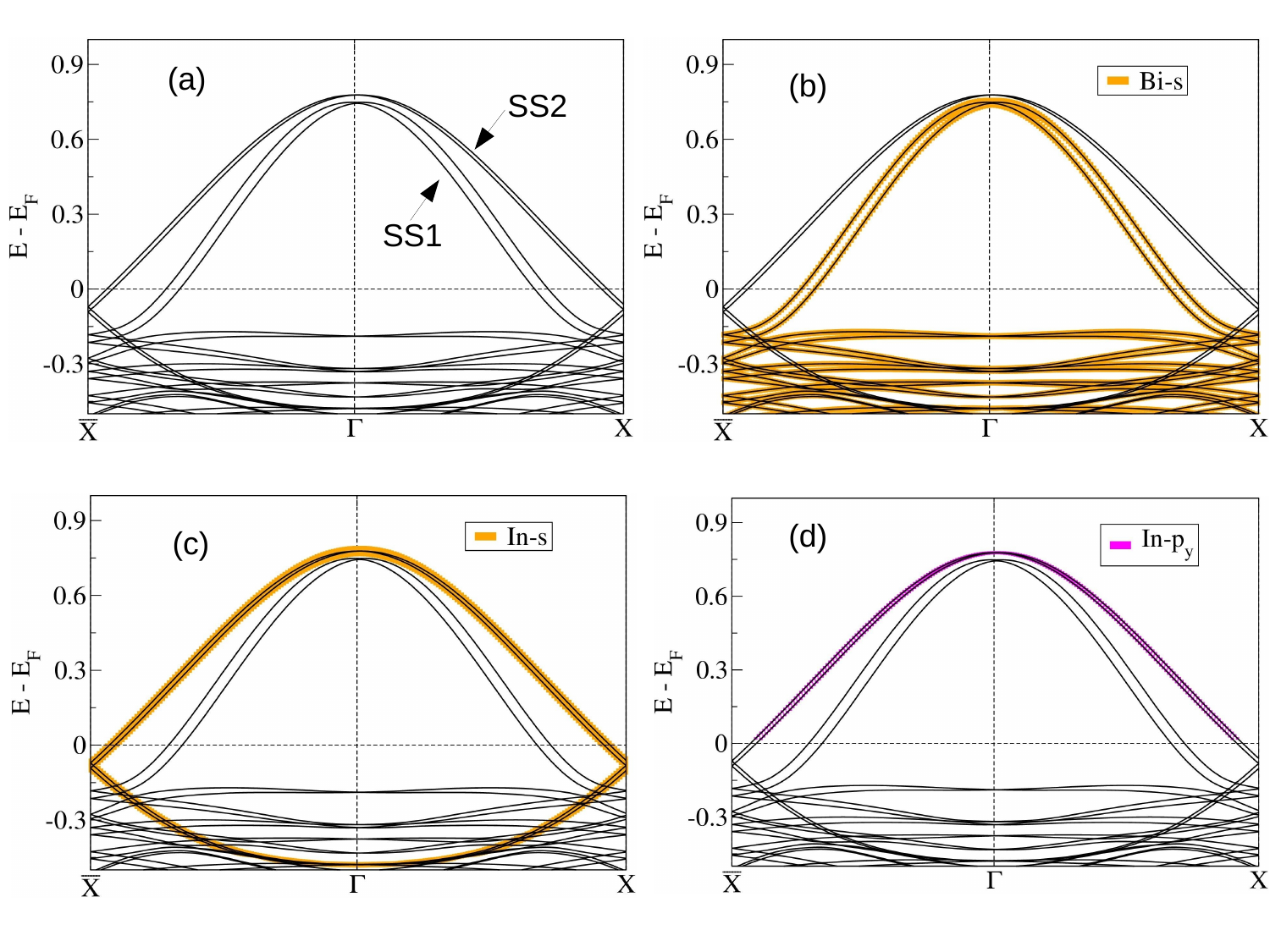}
  \includegraphics[scale=0.5]{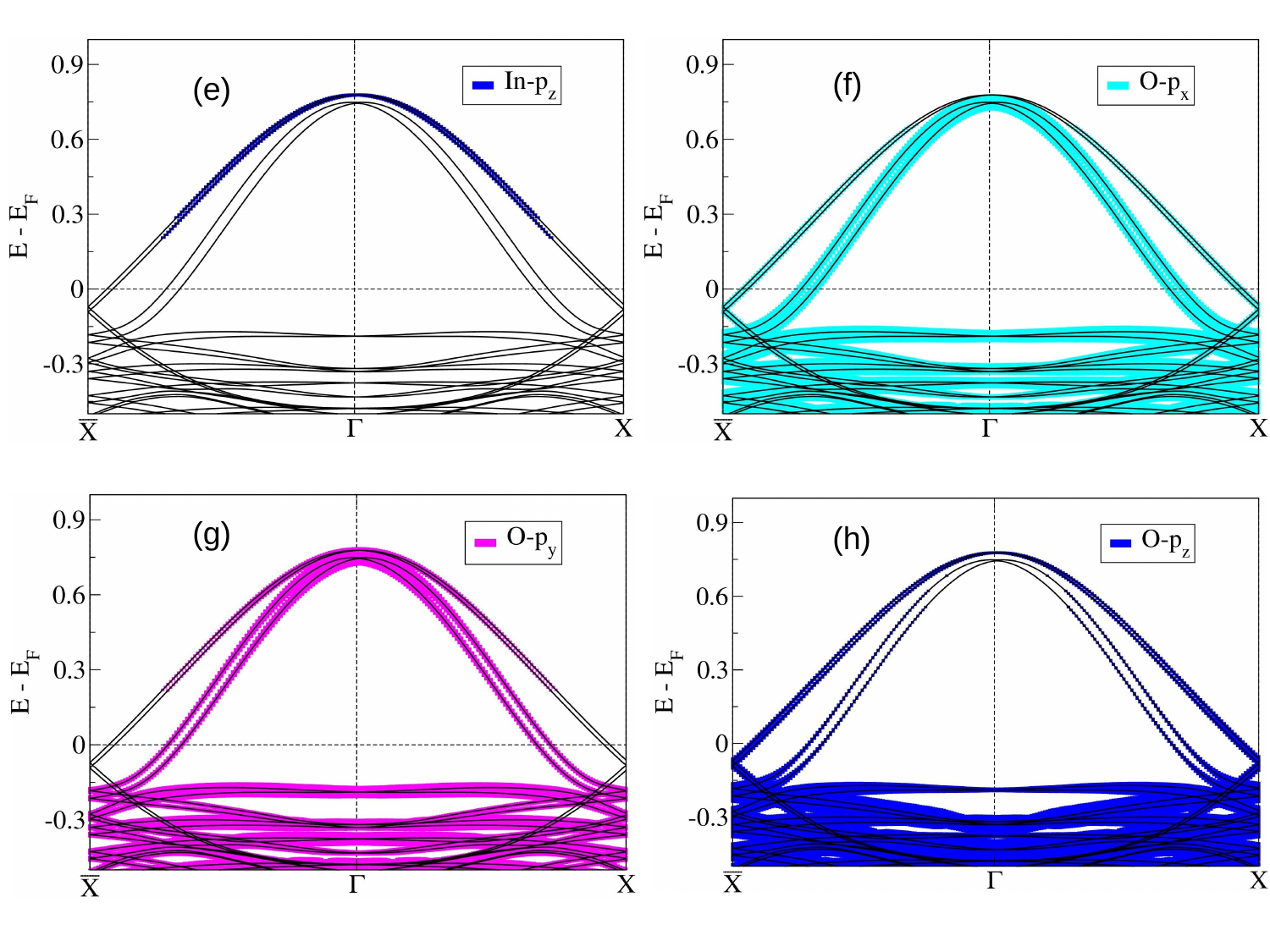}
    \caption{\label{fig:biinfatband}
(a) SS1 and SS2 denote the first and second surface states, respectively.  
(b)–(h) Band dispersion with orbital character projected onto the surface states.  
For SS1, the dominant orbital contributions come from Bi-$s$, O-$p_x$, and O-$p_y$, with a minor contribution from O-$p_z$.  
For SS2, the contributing orbitals include In-$s$, In-$p_y$, In-$p_z$, O-$p_x$, O-$p_y$, and O-$p_z$, among which In-$s$ and O-$p_x$ are the most dominant.
}

\end{figure*}

\subsection{Electronic structure of BiInO$_3$ (001) Surface}

\subsubsection{Band Connectivity Along High-Symmetry Paths $\Gamma$$\rightarrow X$$\rightarrow S $$\rightarrow Y \rightarrow \Gamma$ in Momentum Space}

To understand the evolution of band structure in BiInO$_3$, we explore several high-symmetry paths in momentum space. These paths are chosen based on the symmetries that preserve the wave vector $\vec{k}$ along each segment.

(A) We begin with the path $\Gamma \rightarrow X$, where the wave vector is of the form $(k_x, 0, 0)$, with $k_x$ varying from $0$ to $\pi$, and both $k_y = 0$ and $k_z = 0$ held fixed. Along this path, the wave vector is invariant under the symmetry $\left\{ m_{010} \mid \frac{1}{2}~0~\frac{1}{2} \right\}$. From \cref{eq:14}, the eigenvalues of $\left\{ m_{010} \mid \frac{1}{2}~0~\frac{1}{2} \right\}$ at the $\Gamma$ point ($k_x = 0$) are $(+i, -i)$, whereas at $X$ ($k_x = \pi$), the eigenvalues become $(+1, -1)$. Similarly, at the $-X$ point ($k_x = -\pi$), the eigenvalues are $(-1, +1)$. This change in eigenvalues implies that there are four bands connecting $X \rightarrow \Gamma \rightarrow -X$, and the bands must cross due to internal partner switching between $\Gamma$ and the X (and $-X$) points\cite{Wang2016b,Ma2017} as shown in \cref{fig:xgxband}.

(B) A similar behaviour is observed along the path $S \rightarrow Y$. Here, the wave vector is of the form $(k_x, \pi, 0)$, and $k_x$ is again invariant under the symmetry $\left\{ m_{010} \mid \frac{1}{2}~0~\frac{1}{2} \right\}$, just like in the $\Gamma \rightarrow X$ path. The eigenvalues of $\left\{ m_{010} \mid \frac{1}{2}~0~\frac{1}{2} \right\}$ along this path are $(+i, -i)$ at $-\Tilde{S}$, $(+1, -1)$ at $\Tilde{Y}$, and $(-1, +1)$ at $\Tilde{S}$. This leads to a similar four-band connectivity with symmetry-enforced crossings due to partner switching as shown in \cref{fig:sysbandsmall}.

(C) Next, we consider the path $X \rightarrow S$, traced out by the variable point $(\pi, k_y, 0)$, with $k_y$ varying from $0$ to $\pi$. Along this path, the wave vector is invariant under the antiunitary symmetry $\mathcal{T}\left\{ m_{010} \mid \frac{1}{2}~0~\frac{1}{2} \right\}$. According to \cref{eq:10}, the square of this operator is $\mathcal{T}^2\{m_{010}^2|1~0~0\} = -\identity$ throughout the path. As a result, time-reversal symmetry enforces Kramers degeneracy, and the bands appear in doubly degenerate pairs along the entire $X \rightarrow S$ segment\cite{xie2021kramers}.

(D) Finally, we examine the path $Y \rightarrow \Gamma$, traced by the point $(0, k_y, 0)$, where $k_y$ varies from $0$ to $\pi$. This path is also invariant under the antiunitary symmetry $\mathcal{T}\left\{ m_{010} \mid \frac{1}{2}~0~\frac{1}{2} \right\}$, similar to the $X \rightarrow S$ path. However, here $k_x = 0$ throughout. According to \cref{eq:10}, we find that $\mathcal{T}^2\{m_{010}^2|1~0~0\} = \identity$, meaning that time-reversal symmetry alone does not enforce degeneracy. Therefore, along this path, the bands split in the range $0 \leq k_y \leq \pi$.

\subsubsection{$\vec{k} \cdot \vec{p}$  Hamiltonian at (001)-Surface}
The energy band splitting and spin texture of BiInO$_3$ near the $\Gamma$ point can be effectively described using a $\vec{k} \cdot \vec{p}$ Hamiltonian that incorporates all relevant symmetries at this high-symmetry point. The point group at $\Gamma$ is $\mathcal{C}_{2v}$, which includes a twofold rotation about the $z$-axis, $\mathcal{C}_{2z} = \left\{ 2_{001} \mid 0~0~0 \right\}$, as well as mirror reflections about the $xz$- and $yz$-planes, denoted by $\mathcal{M}_y = \left\{ m_{010} \mid 0~0~0 \right\}$ and $\mathcal{M}_x = \left\{ m_{100} \mid 0~0~0 \right\}$, respectively. Since BiInO$_3$ is non-magnetic, time-reversal symmetry $\mathcal{T}$ is preserved. Under time-reversal, the spin operator $\vec{\sigma}$ and the crystal momentum $\vec{k}$ transform as $\vec{\sigma} \rightarrow -\vec{\sigma}$ and $\vec{k} \rightarrow -\vec{k}$, respectively. The transformation properties of both spin and momentum vectors under these symmetry operations are summarized in Table~\ref{tab:ch3tabtrans}. Based on the method of invariants, the symmetry-allowed linear-order $\vec{k} \cdot \vec{p}$ Hamiltonian for bulk BiInO$_3$ near the $\Gamma$ point can be written as follows:

\begin{equation}
    \mathcal{H}=\mathcal{H}_o + \mathcal{H}_{so}\label{eq:11}
\end{equation}
Where,
\begin{equation}
    \mathcal{H}_{so}= \alpha_1k_x\sigma_y+\alpha_2k_y\sigma_x \label{eq:12}
\end{equation}
\begin{equation}
    \mathcal{H}_o= \dfrac{\hbar^2k_x^2}{2m_x}+\dfrac{\hbar^2k_y^2}{2m_y}\label{eq:13}
\end{equation}
However, at the 001 surfaces, translation symmetry along the z-axis gets terminated, resulting in structure inversion asymmetry(SIA). At the surface $\left\{ m_{010} \mid \frac{1}{2}~0~\frac{1}{2} \right\}$ is only the symmetry present at the (001)-surface of BiInO$_3$. Under the symmetry $\left\{ m_{010} \mid \frac{1}{2}~0~\frac{1}{2} \right\}$ component of $\vec{k}$ transform as $(k_x,k_y,k_z)\rightarrow (k_x,-k_y,k_z$) and Pauli spin transform as $(\sigma_x,\sigma_y,\sigma_z)\rightarrow(-\sigma_x,\sigma_y,-\sigma_z)$. Similarly, under Time reversal symmetry $\mathcal{T}$ component of $\vec{k}$ transform as $(k_x,k_y,k_z)\rightarrow (-k_x,-k_y,-k_z$) and Pauli spin transform as $(\sigma_x,\sigma_y,\sigma_z)\rightarrow(-\sigma_x,-\sigma_y,-\sigma_z)$. The transformation rule of $\vec(k)$ and $\sigma$ is summarized in \cref{tab:ch3tabtrans}. $k_x\sigma_y$  and $k_y\sigma_x$ will be invariant under both symmetry $\left\{ m_{010} \mid 0~0~0 \right\}$ and $\mathcal{T}$ and contribute to the terms for Hamiltonian due to Spin-Orbit Interaction at the surface.
If we consider $\alpha_1=\alpha_2=\alpha_D$ in \cref{eq:12} we will get Hamiltonian due to bulk inversion asymmetry called Dresselhaus Hamiltonian as,
\begin{equation}
     \mathcal{H}_{BIA}= \alpha_D (k_x\sigma_y+k_y\sigma_x)\label{eq:14}
\end{equation}

\begin{table*}[t]
\centering
\caption{Transformation rules for wave vector $\vec{k}$ and spin ($\sigma$) under the 2D point-group `m' symmetry operations at the $\Gamma(0, 0, 0)$ point in the Brillouin zone of BiInO$_3$. The wave vector $\vec{k}$ is referenced for the high symmetry point. $\hat{\mathcal{K}}$ denotes the complex conjugation operator.}
\label{tab:ch3tabtrans}
\begin{tabular}{>{\centering\arraybackslash}m{4cm} >{\centering\arraybackslash}m{3cm} >{\centering\arraybackslash}m{4cm} >{\centering\arraybackslash}m{3cm}}
\hline \hline
\textrm{Symmetry Operator} & \textrm{Wave vector ($\vec{k}$)} & \textrm{Spin ($\sigma$)} & \textrm{Invariants} \\
\hline
$\mathcal{T}= i\sigma{_y} \hat{\mathcal{K}}$ & ($-k_x,-k_y$) & ($-\sigma_x,-\sigma_y,-\sigma_z$) & $k_i\sigma_i, \quad i=x,y,z$ \\
$\left\{ m_{010} \mid \frac{1}{2}~0~\frac{1}{2} \right\}=i\sigma_y$ & ($k_x,-k_y$) & ($-\sigma_x,\sigma_y,-\sigma_z$) & $k_x\sigma_y, k_y\sigma_x$ \\
\hline \hline
\end{tabular}
\end{table*}

If we consider $\alpha_1=-\alpha_2= \alpha_R$ in \cref{eq:12} we will get Hamiltonian due to surface inversion asymmetry called Rashba Hamiltonian as,
\begin{equation}
    \mathcal{H}_{SIA}=\alpha_R(k_x\sigma_y-k_y\sigma_x) \label{eq:15}
\end{equation}
The Hamiltonian for the energy band dispersion at the (001)-surface of BiInO$_3$ having contribution coming from both bulk and surface asymmetry  will be,
\begin{equation}
    \mathcal{H}= \mathcal{H}_o + \mathcal{H}_{(BIA+SIA)} \label{eq:16}
\end{equation}
\begin{equation} \label{biasia}
    \mathcal{H}_{(BIA+SIA)}=\alpha_D(k_x\sigma_y+k_y\sigma_x)+\alpha_R(k_x\sigma_y-k_y\sigma_x) 
\end{equation}
Diagonalizing \cref{biasia} Energy eigenvalue we get,
\begin{equation}
\varepsilon^{\pm}(k)=\dfrac{k_x^2}{2m_x}+\dfrac{k_y^2}{2m_y}\pm \sqrt{(\alpha_D-\alpha_R)^2k_y^2+(\alpha_D+\alpha_R)^2k_x^2}    \label{eq:18}
\end{equation}
The eigenstates of \cref{biasia} in spinor form is,
\begin{equation}
    \psi(\vec{k})^{\pm}(k)=\frac{1}{\sqrt{2}}(\pm\xi \ket{\uparrow}+\ket{\downarrow})  \label{eq:19}
\end{equation}
\begin{equation}
    \xi =\dfrac{(\alpha_D-\alpha_R)k_y-i(\alpha_D+\alpha_R)k_x}{\sqrt{(\alpha_D-\alpha_R)^2k_y^2+(\alpha_D+\alpha_R)^2k_x^2}}    \label{eq:20}
\end{equation}

$\ket{\uparrow}$ and $\ket{\downarrow}$ are the spinors for electrons spinning up and down.

The expectation value of spin operator is $s^{\pm}=\frac{1}{2}\bra{\psi(\vec{k})^{\pm}(k)}\sigma\ket{\psi(\vec{k})^{\pm}(k)}$ results,
\begin{equation}
   \langle s_x\rangle^{\pm}=\pm \frac{1}{2}\left(\frac{(\alpha_D-\alpha_R)k_y}{\sqrt{(\alpha_D-\alpha_R)^2k_y^2+(\alpha_D+\alpha_R)^2k_x^2}} \right)   \label{eq:21}
\end{equation}
\begin{equation}
  \langle s_y\rangle^{\pm}=\pm \frac{1}{2}\left(\frac{(\alpha_D+\alpha_R)k_x}{\sqrt{(\alpha_D-\alpha_R)^2k_y^2+(\alpha_D+\alpha_R)^2k_x^2}} \right)     \label{eq:22}
\end{equation}
\begin{equation}
  \langle s_z\rangle^{\pm}= 0  \label{eq:23}
\end{equation}
 The competition between Rashba and Dresselhaus SOI gives rise to different types of energy bands and spin textures given in \cref{fig:RD_image}. If $\alpha_D=0$ and a non-zero value of $\alpha_R$ in \cref{eq:18}, we will get only Rashba SOI. However, we will get only Dresselhaus SOI if $\alpha_R=0$ and non-zero $\alpha_D$ in \cref{eq:18}. If both $\alpha_D$ and $\alpha_R$ are non-zero, we will get a mixture of Rashba and Dresselhaus SOI. In a particular case, if $\alpha_D=\alpha_R$, we will get PST.

\subsection{Surface States and Spin Texture on the (001) Surface of BiInO\textsubscript{3}}

At the (001) surface of BiInO$_3$, two sets of conducting bands are observed to split along the momentum direction. These are denoted as Surface State 1 (SS1) and Surface State 2 (SS2), as shown in \cref{fig:biinfatband}(a). The orbital contributions to these bands play a crucial role in determining the spin direction, and this can be understood through a simple symmetry analysis based on Neumann’s principle. According to Neumann’s principle, a physical property described by a tensor must remain invariant under any symmetry operation of the point group associated with a given high-symmetry point in the crystal\cite{nye1985physical,sirotin1982fundamentals}. Therefore, the transformation behaviour of orbitals and their associated spin components must follow the symmetries present at that point\cite{PhysRevB.103.014105}.

To explore the orbital influence on spin orientation, we calculated the orbital-projected energy bands shown in \cref{fig:biinfatband}. For SS2, the contributing orbitals are In-$s$, In-$p_y$, In-$p_z$, O-$p_x$, O-$p_y$, and O-$p_z$, with dominant contributions from O-$p_x$ and In-$s$. Under the mirror symmetry $\left\{ m_{010} \mid 0~0~0 \right\}$, the $p_x$ orbital transforms as $p_x \rightarrow p_x$, and the spin component $\sigma_y$ transforms similarly as $\sigma_y \rightarrow \sigma_y$. In-$s$ orbitals, being spherically symmetric, remain invariant under all symmetry operations. Because both O-$p_x$ and $\sigma_y$ transform identically under $\left\{ m_{010} \mid 0~0~0 \right\}$, the spin orientation for SS2 aligns along the $y$-direction.

For SS1, the contributing orbitals are Bi-$s$, O-$p_x$, O-$p_y$, and O-$p_z$, where Bi-$s$ and O-$p_z$ have minimal contributions, and O-$p_x$ and O-$p_y$ dominate. Under $\left\{ m_{010} \mid 0~0~0 \right\}$ symmetry, the $p_x$ orbital transforms as $p_x \rightarrow p_x$, while $p_y \rightarrow -p_y$. The spin components transform accordingly: $\sigma_x \rightarrow -\sigma_x$ and $\sigma_y \rightarrow \sigma_y$. Since O-$p_x$ and O-$p_y$ contribute nearly equally to SS1, the resulting spin orientation is a combination of both $k_x$ and $k_y$ directions at different $k$-points. This results in a Rashba-like spin texture, as confirmed by our DFT calculations.

To further analyse the spin structure, we fit a $\vec{k} \cdot \vec{p}$ model to the DFT band structure. For SS1, using eigenstates from \cref{eq:18}, we found spin–orbit interaction (SOI) constants $\alpha_R = 0.20$~eV\AA, $\alpha_D = 0$, and isotropic effective masses $m_x = m_y = -0.053~m_e$, where $m_e$ is the electron rest mass. Using these values, we plotted the energy bands and spin textures, which show spins tangential to the circular contours of constant energy. This confirms that SS1 exhibits Rashba-like spin splitting with a relatively light isotropic effective mass. The agreement between this analytical result and DFT calculations validates the orbital symmetry analysis discussed earlier.

Similarly, we fit the SS2 surface band using the $\vec{k} \cdot \vec{p}$ model and DFT data. We found equal Rashba and Dresselhaus SOI constants, $\alpha_R = \alpha_D = 0.0205$~eV\AA, with effective masses $m_x = -0.082~m_e$ and $m_y = -0.355~m_e$. The spin texture plotted with these parameters reveals a unidirectional spin orientation along the $y$-axis. This momentum-independent spin configuration is consistent with a Persistent Spin Texture (PST), as expected from the equal contributions of Rashba and Dresselhaus interactions. The anisotropy in effective mass between $x$ and $y$ directions further indicates strong directional dependence in the band dispersion for SS2.

\section{\label{sec:conc}Conclusion} This study delves into the role of symmetry-enforced band crossing on the surface band of BIO [001] surface. We have highlighted how symmetry plays a role at the [001] surface in determining the Hourglass fermion in the surface state. Using density functional theory (DFT), symmetry analysis, and the $\vec{k} \cdot \vec{p}$ model, the research uncovers the fascinating spin-momentum locking at the surface band due to the dangling bond. The DFT band fitted with the $\vec{k}\cdot \vec{p}$ model reveals that SS1 demonstrates a Rashba-like spin texture, dominated by Bi-s, O-p$_x$, and O-p$_y$ orbitals, with spin alignment tangential to the constant energy contours. In contrast, SS2 exhibits equal contributions from Rashba and Dresselhaus spin-orbit interactions (SOI), resulting in a Persistent Spin Texture (PST) with uni-directional spin alignment along the k$_y$-axis. These findings not only deepen the understanding of surface electronic behaviour in BIO but also position this material as a promising candidate for spintronic devices. The observed spin textures, particularly PST, offer potential for high-efficiency spin-based applications by reducing spin relaxation effects\cite{Dyakonov2017}.

\begin{acknowledgments}
RK acknowledges UGC, India, for a research fellowship through grant number 1500/(CSIR-UGC NET JUNE 2017). NG acknowledges financial support from SERB, India, through grant number CRG/2021/005320. The use of high-performance computing facilities at IISER Bhopal and IIT Hyderabad under the National Computing Mission (NSM) is greatly acknowledged.
\end{acknowledgments}

   


\bibliography{library1}

\begin{thebibliography}{43}%
\makeatletter
\providecommand \@ifxundefined [1]{%
 \@ifx{#1\undefined}
}%
\providecommand \@ifnum [1]{%
 \ifnum #1\expandafter \@firstoftwo
 \else \expandafter \@secondoftwo
 \fi
}%
\providecommand \@ifx [1]{%
 \ifx #1\expandafter \@firstoftwo
 \else \expandafter \@secondoftwo
 \fi
}%
\providecommand \natexlab [1]{#1}%
\providecommand \enquote  [1]{``#1''}%
\providecommand \bibnamefont  [1]{#1}%
\providecommand \bibfnamefont [1]{#1}%
\providecommand \citenamefont [1]{#1}%
\providecommand \href@noop [0]{\@secondoftwo}%
\providecommand \href [0]{\begingroup \@sanitize@url \@href}%
\providecommand \@href[1]{\@@startlink{#1}\@@href}%
\providecommand \@@href[1]{\endgroup#1\@@endlink}%
\providecommand \@sanitize@url [0]{\catcode `\\12\catcode `\$12\catcode `\&12\catcode `\#12\catcode `\^12\catcode `\_12\catcode `\%12\relax}%
\providecommand \@@startlink[1]{}%
\providecommand \@@endlink[0]{}%
\providecommand \url  [0]{\begingroup\@sanitize@url \@url }%
\providecommand \@url [1]{\endgroup\@href {#1}{\urlprefix }}%
\providecommand \urlprefix  [0]{URL }%
\providecommand \Eprint [0]{\href }%
\providecommand \doibase [0]{https://doi.org/}%
\providecommand \selectlanguage [0]{\@gobble}%
\providecommand \bibinfo  [0]{\@secondoftwo}%
\providecommand \bibfield  [0]{\@secondoftwo}%
\providecommand \translation [1]{[#1]}%
\providecommand \BibitemOpen [0]{}%
\providecommand \bibitemStop [0]{}%
\providecommand \bibitemNoStop [0]{.\EOS\space}%
\providecommand \EOS [0]{\spacefactor3000\relax}%
\providecommand \BibitemShut  [1]{\csname bibitem#1\endcsname}%
\let\auto@bib@innerbib\@empty
\bibitem [{\citenamefont {Kane}\ and\ \citenamefont {Mele}(2005)}]{Kane2005}%
  \BibitemOpen
  \bibfield  {author} {\bibinfo {author} {\bibfnamefont {C.~L.}\ \bibnamefont {Kane}}\ and\ \bibinfo {author} {\bibfnamefont {E.~J.}\ \bibnamefont {Mele}},\ }\bibfield  {title} {\bibinfo {title} {{Quantum Spin hall effect in graphene}},\ }\href {https://doi.org/10.1103/PhysRevLett.95.226801} {\bibfield  {journal} {\bibinfo  {journal} {Physical Review Letters}\ }\textbf {\bibinfo {volume} {95}},\ \bibinfo {pages} {1} (\bibinfo {year} {2005})}\BibitemShut {NoStop}%
\bibitem [{\citenamefont {Moore}\ and\ \citenamefont {Balents}(2007)}]{Moore2007}%
  \BibitemOpen
  \bibfield  {author} {\bibinfo {author} {\bibfnamefont {J.~E.}\ \bibnamefont {Moore}}\ and\ \bibinfo {author} {\bibfnamefont {L.}~\bibnamefont {Balents}},\ }\bibfield  {title} {\bibinfo {title} {{Topological invariants of time-reversal-invariant band structures}},\ }\href {https://doi.org/10.1103/PhysRevB.75.121306} {\bibfield  {journal} {\bibinfo  {journal} {Physical Review B - Condensed Matter and Materials Physics}\ }\textbf {\bibinfo {volume} {75}},\ \bibinfo {pages} {3} (\bibinfo {year} {2007})}\BibitemShut {NoStop}%
\bibitem [{\citenamefont {Fu}\ and\ \citenamefont {Kane}(2006)}]{Fu2006}%
  \BibitemOpen
  \bibfield  {author} {\bibinfo {author} {\bibfnamefont {L.}~\bibnamefont {Fu}}\ and\ \bibinfo {author} {\bibfnamefont {C.~L.}\ \bibnamefont {Kane}},\ }\bibfield  {title} {\bibinfo {title} {{Time reversal polarization and a Z2 adiabatic spin pump}},\ }\href {https://doi.org/10.1103/PhysRevB.74.195312} {\bibfield  {journal} {\bibinfo  {journal} {Physical Review B - Condensed Matter and Materials Physics}\ }\textbf {\bibinfo {volume} {74}},\ \bibinfo {pages} {1} (\bibinfo {year} {2006})}\BibitemShut {NoStop}%
\bibitem [{\citenamefont {Chiu}\ \emph {et~al.}(2016)\citenamefont {Chiu}, \citenamefont {Teo}, \citenamefont {Schnyder},\ and\ \citenamefont {Ryu}}]{Chiu2016}%
  \BibitemOpen
  \bibfield  {author} {\bibinfo {author} {\bibfnamefont {C.~K.}\ \bibnamefont {Chiu}}, \bibinfo {author} {\bibfnamefont {J.~C.}\ \bibnamefont {Teo}}, \bibinfo {author} {\bibfnamefont {A.~P.}\ \bibnamefont {Schnyder}},\ and\ \bibinfo {author} {\bibfnamefont {S.}~\bibnamefont {Ryu}},\ }\bibfield  {title} {\bibinfo {title} {{Classification of topological quantum matter with symmetries}},\ }\href {https://doi.org/10.1103/RevModPhys.88.035005} {\bibfield  {journal} {\bibinfo  {journal} {Reviews of Modern Physics}\ }\textbf {\bibinfo {volume} {88}},\ \bibinfo {pages} {1} (\bibinfo {year} {2016})}\BibitemShut {NoStop}%
\bibitem [{\citenamefont {Fu}\ and\ \citenamefont {Kane}(2007)}]{Fu2007}%
  \BibitemOpen
  \bibfield  {author} {\bibinfo {author} {\bibfnamefont {L.}~\bibnamefont {Fu}}\ and\ \bibinfo {author} {\bibfnamefont {C.~L.}\ \bibnamefont {Kane}},\ }\bibfield  {title} {\bibinfo {title} {{Topological insulators with inversion symmetry}},\ }\href {https://doi.org/10.1103/PhysRevB.76.045302} {\bibfield  {journal} {\bibinfo  {journal} {Physical Review B - Condensed Matter and Materials Physics}\ }\textbf {\bibinfo {volume} {76}},\ \bibinfo {pages} {1} (\bibinfo {year} {2007})}\BibitemShut {NoStop}%
\bibitem [{\citenamefont {Achal}\ \emph {et~al.}(2018)\citenamefont {Achal}, \citenamefont {Rashidi}, \citenamefont {Croshaw}, \citenamefont {Churchill}, \citenamefont {Taucer}, \citenamefont {Huff}, \citenamefont {Cloutier}, \citenamefont {Pitters},\ and\ \citenamefont {Wolkow}}]{achal2018lithography}%
  \BibitemOpen
  \bibfield  {author} {\bibinfo {author} {\bibfnamefont {R.}~\bibnamefont {Achal}}, \bibinfo {author} {\bibfnamefont {M.}~\bibnamefont {Rashidi}}, \bibinfo {author} {\bibfnamefont {J.}~\bibnamefont {Croshaw}}, \bibinfo {author} {\bibfnamefont {D.}~\bibnamefont {Churchill}}, \bibinfo {author} {\bibfnamefont {M.}~\bibnamefont {Taucer}}, \bibinfo {author} {\bibfnamefont {T.}~\bibnamefont {Huff}}, \bibinfo {author} {\bibfnamefont {M.}~\bibnamefont {Cloutier}}, \bibinfo {author} {\bibfnamefont {J.}~\bibnamefont {Pitters}},\ and\ \bibinfo {author} {\bibfnamefont {R.~A.}\ \bibnamefont {Wolkow}},\ }\bibfield  {title} {\bibinfo {title} {Lithography for robust and editable atomic-scale silicon devices and memories},\ }\href {https://doi.org/10.1038/s41467-018-05171-y} {\bibfield  {journal} {\bibinfo  {journal} {Nature Communications}\ }\textbf {\bibinfo {volume} {9}},\ \bibinfo {pages} {2778} (\bibinfo {year} {2018})}\BibitemShut {NoStop}%
\bibitem [{\citenamefont {Berthe}\ \emph {et~al.}(2008)\citenamefont {Berthe}, \citenamefont {Stiufiuc}, \citenamefont {Grandidier}, \citenamefont {Deresmes}, \citenamefont {Delerue},\ and\ \citenamefont {Stiévenard}}]{berthe2008probing}%
  \BibitemOpen
  \bibfield  {author} {\bibinfo {author} {\bibfnamefont {M.}~\bibnamefont {Berthe}}, \bibinfo {author} {\bibfnamefont {R.}~\bibnamefont {Stiufiuc}}, \bibinfo {author} {\bibfnamefont {B.}~\bibnamefont {Grandidier}}, \bibinfo {author} {\bibfnamefont {D.}~\bibnamefont {Deresmes}}, \bibinfo {author} {\bibfnamefont {C.}~\bibnamefont {Delerue}},\ and\ \bibinfo {author} {\bibfnamefont {D.}~\bibnamefont {Stiévenard}},\ }\bibfield  {title} {\bibinfo {title} {Probing the carrier capture rate of a single quantum level},\ }\href {https://doi.org/10.1126/science.1151186} {\bibfield  {journal} {\bibinfo  {journal} {Science}\ }\textbf {\bibinfo {volume} {319}},\ \bibinfo {pages} {436} (\bibinfo {year} {2008})},\ \bibinfo {note} {epub 2007 Dec 13}\BibitemShut {NoStop}%
\bibitem [{\citenamefont {Boland}(1993)}]{boland1993manipulating}%
  \BibitemOpen
  \bibfield  {author} {\bibinfo {author} {\bibfnamefont {J.~J.}\ \bibnamefont {Boland}},\ }\bibfield  {title} {\bibinfo {title} {Manipulating chlorine atom bonding on the si (100)-(2$\times$ 1) surface with the stm},\ }\href@noop {} {\bibfield  {journal} {\bibinfo  {journal} {Science}\ }\textbf {\bibinfo {volume} {262}},\ \bibinfo {pages} {1703} (\bibinfo {year} {1993})}\BibitemShut {NoStop}%
\bibitem [{\citenamefont {Lin}\ \emph {et~al.}(2013)\citenamefont {Lin}, \citenamefont {Das}, \citenamefont {Okada}, \citenamefont {Boyer}, \citenamefont {Wise}, \citenamefont {Tomasik}, \citenamefont {Zhen}, \citenamefont {Hudson}, \citenamefont {Zhou}, \citenamefont {Madhavan}, \citenamefont {Ren}, \citenamefont {Ikuta},\ and\ \citenamefont {Bansil}}]{lin2013topological}%
  \BibitemOpen
  \bibfield  {author} {\bibinfo {author} {\bibfnamefont {H.}~\bibnamefont {Lin}}, \bibinfo {author} {\bibfnamefont {T.}~\bibnamefont {Das}}, \bibinfo {author} {\bibfnamefont {Y.}~\bibnamefont {Okada}}, \bibinfo {author} {\bibfnamefont {M.~C.}\ \bibnamefont {Boyer}}, \bibinfo {author} {\bibfnamefont {W.~D.}\ \bibnamefont {Wise}}, \bibinfo {author} {\bibfnamefont {M.}~\bibnamefont {Tomasik}}, \bibinfo {author} {\bibfnamefont {B.}~\bibnamefont {Zhen}}, \bibinfo {author} {\bibfnamefont {E.~W.}\ \bibnamefont {Hudson}}, \bibinfo {author} {\bibfnamefont {W.}~\bibnamefont {Zhou}}, \bibinfo {author} {\bibfnamefont {V.}~\bibnamefont {Madhavan}}, \bibinfo {author} {\bibfnamefont {C.-Y.}\ \bibnamefont {Ren}}, \bibinfo {author} {\bibfnamefont {H.}~\bibnamefont {Ikuta}},\ and\ \bibinfo {author} {\bibfnamefont {A.}~\bibnamefont {Bansil}},\ }\bibfield  {title} {\bibinfo {title} {Topological dangling bonds with large spin splitting and enhanced spin polarization on the surfaces of bi$_2$se$_3$},\ }\href
  {https://doi.org/10.1021/nl304099x} {\bibfield  {journal} {\bibinfo  {journal} {Nano Letters}\ }\textbf {\bibinfo {volume} {13}},\ \bibinfo {pages} {1915} (\bibinfo {year} {2013})}\BibitemShut {NoStop}%
\bibitem [{\citenamefont {Wang}\ \emph {et~al.}(2016)\citenamefont {Wang}, \citenamefont {Alexandradinata}, \citenamefont {Cava},\ and\ \citenamefont {Bernevig}}]{Wang2016b}%
  \BibitemOpen
  \bibfield  {author} {\bibinfo {author} {\bibfnamefont {Z.}~\bibnamefont {Wang}}, \bibinfo {author} {\bibfnamefont {A.}~\bibnamefont {Alexandradinata}}, \bibinfo {author} {\bibfnamefont {R.~J.}\ \bibnamefont {Cava}},\ and\ \bibinfo {author} {\bibfnamefont {B.~A.}\ \bibnamefont {Bernevig}},\ }\bibfield  {title} {\bibinfo {title} {{Hourglass fermions}},\ }\href {https://doi.org/10.1038/nature17410} {\bibfield  {journal} {\bibinfo  {journal} {Nature}\ }\textbf {\bibinfo {volume} {532}},\ \bibinfo {pages} {189} (\bibinfo {year} {2016})}\BibitemShut {NoStop}%
\bibitem [{\citenamefont {Ma}\ \emph {et~al.}(2017)\citenamefont {Ma}, \citenamefont {Yi}, \citenamefont {Lv}, \citenamefont {Wang}, \citenamefont {Nie}, \citenamefont {Wang}, \citenamefont {Kong}, \citenamefont {Huang}, \citenamefont {Richard}, \citenamefont {Zhang}, \citenamefont {Yaji}, \citenamefont {Kuroda}, \citenamefont {Shin}, \citenamefont {Weng}, \citenamefont {Bernevig}, \citenamefont {Shi}, \citenamefont {Qian},\ and\ \citenamefont {Ding}}]{Ma2017}%
  \BibitemOpen
  \bibfield  {author} {\bibinfo {author} {\bibfnamefont {J.}~\bibnamefont {Ma}}, \bibinfo {author} {\bibfnamefont {C.}~\bibnamefont {Yi}}, \bibinfo {author} {\bibfnamefont {B.}~\bibnamefont {Lv}}, \bibinfo {author} {\bibfnamefont {Z.}~\bibnamefont {Wang}}, \bibinfo {author} {\bibfnamefont {S.}~\bibnamefont {Nie}}, \bibinfo {author} {\bibfnamefont {L.}~\bibnamefont {Wang}}, \bibinfo {author} {\bibfnamefont {L.}~\bibnamefont {Kong}}, \bibinfo {author} {\bibfnamefont {Y.}~\bibnamefont {Huang}}, \bibinfo {author} {\bibfnamefont {P.}~\bibnamefont {Richard}}, \bibinfo {author} {\bibfnamefont {P.}~\bibnamefont {Zhang}}, \bibinfo {author} {\bibfnamefont {K.}~\bibnamefont {Yaji}}, \bibinfo {author} {\bibfnamefont {K.}~\bibnamefont {Kuroda}}, \bibinfo {author} {\bibfnamefont {S.}~\bibnamefont {Shin}}, \bibinfo {author} {\bibfnamefont {H.}~\bibnamefont {Weng}}, \bibinfo {author} {\bibfnamefont {B.~A.}\ \bibnamefont {Bernevig}}, \bibinfo {author} {\bibfnamefont {Y.}~\bibnamefont {Shi}}, \bibinfo {author} {\bibfnamefont
  {T.}~\bibnamefont {Qian}},\ and\ \bibinfo {author} {\bibfnamefont {H.}~\bibnamefont {Ding}},\ }\bibfield  {title} {\bibinfo {title} {{Experimental evidence of hourglass fermion in the candidate nonsymmorphic topological insulator KHgSb}},\ }\href {https://doi.org/10.1126/sciadv.1602415} {\bibfield  {journal} {\bibinfo  {journal} {Science Advances}\ }\textbf {\bibinfo {volume} {3}},\ \bibinfo {pages} {2} (\bibinfo {year} {2017})}\BibitemShut {NoStop}%
\bibitem [{\citenamefont {Wang}\ \emph {et~al.}(2017{\natexlab{a}})\citenamefont {Wang}, \citenamefont {Liu}, \citenamefont {Yu}, \citenamefont {Sheng},\ and\ \citenamefont {Yang}}]{Wang2017}%
  \BibitemOpen
  \bibfield  {author} {\bibinfo {author} {\bibfnamefont {S.~S.}\ \bibnamefont {Wang}}, \bibinfo {author} {\bibfnamefont {Y.}~\bibnamefont {Liu}}, \bibinfo {author} {\bibfnamefont {Z.~M.}\ \bibnamefont {Yu}}, \bibinfo {author} {\bibfnamefont {X.~L.}\ \bibnamefont {Sheng}},\ and\ \bibinfo {author} {\bibfnamefont {S.~A.}\ \bibnamefont {Yang}},\ }\bibfield  {title} {\bibinfo {title} {{Hourglass Dirac chain metal in rhenium dioxide}},\ }\href {https://doi.org/10.1038/s41467-017-01986-3} {\bibfield  {journal} {\bibinfo  {journal} {Nature Communications}\ }\textbf {\bibinfo {volume} {8}},\ \bibinfo {pages} {1} (\bibinfo {year} {2017}{\natexlab{a}})}\BibitemShut {NoStop}%
\bibitem [{\citenamefont {Wang}\ \emph {et~al.}(2019)\citenamefont {Wang}, \citenamefont {Liu},\ and\ \citenamefont {Zhu}}]{Wang2019}%
  \BibitemOpen
  \bibfield  {author} {\bibinfo {author} {\bibfnamefont {Z.~F.}\ \bibnamefont {Wang}}, \bibinfo {author} {\bibfnamefont {B.}~\bibnamefont {Liu}},\ and\ \bibinfo {author} {\bibfnamefont {W.}~\bibnamefont {Zhu}},\ }\bibfield  {title} {\bibinfo {title} {{Hourglass Fermion in Two-Dimensional Material}},\ }\href {https://doi.org/10.1103/PhysRevLett.123.126403} {\bibfield  {journal} {\bibinfo  {journal} {Physical Review Letters}\ }\textbf {\bibinfo {volume} {123}},\ \bibinfo {pages} {126403} (\bibinfo {year} {2019})}\BibitemShut {NoStop}%
\bibitem [{\citenamefont {Wang}\ \emph {et~al.}(2017{\natexlab{b}})\citenamefont {Wang}, \citenamefont {Jian},\ and\ \citenamefont {Yao}}]{PhysRevB.96.075110}%
  \BibitemOpen
  \bibfield  {author} {\bibinfo {author} {\bibfnamefont {L.}~\bibnamefont {Wang}}, \bibinfo {author} {\bibfnamefont {S.-K.}\ \bibnamefont {Jian}},\ and\ \bibinfo {author} {\bibfnamefont {H.}~\bibnamefont {Yao}},\ }\bibfield  {title} {\bibinfo {title} {Hourglass semimetals with nonsymmorphic symmetries in three dimensions},\ }\href {https://doi.org/10.1103/PhysRevB.96.075110} {\bibfield  {journal} {\bibinfo  {journal} {Phys. Rev. B}\ }\textbf {\bibinfo {volume} {96}},\ \bibinfo {pages} {075110} (\bibinfo {year} {2017}{\natexlab{b}})}\BibitemShut {NoStop}%
\bibitem [{\citenamefont {Zhao}\ and\ \citenamefont {Schnyder}(2016)}]{PhysRevB.94.195109}%
  \BibitemOpen
  \bibfield  {author} {\bibinfo {author} {\bibfnamefont {Y.~X.}\ \bibnamefont {Zhao}}\ and\ \bibinfo {author} {\bibfnamefont {A.~P.}\ \bibnamefont {Schnyder}},\ }\bibfield  {title} {\bibinfo {title} {Nonsymmorphic symmetry-required band crossings in topological semimetals},\ }\href {https://doi.org/10.1103/PhysRevB.94.195109} {\bibfield  {journal} {\bibinfo  {journal} {Phys. Rev. B}\ }\textbf {\bibinfo {volume} {94}},\ \bibinfo {pages} {195109} (\bibinfo {year} {2016})}\BibitemShut {NoStop}%
\bibitem [{\citenamefont {Liu}\ \emph {et~al.}(2023)\citenamefont {Liu}, \citenamefont {Huang}, \citenamefont {Sankar}, \citenamefont {Hlevyack}, \citenamefont {Su}, \citenamefont {Weng}, \citenamefont {Lin}, \citenamefont {Chen}, \citenamefont {Cheng}, \citenamefont {Denlinger}, \citenamefont {Mo}, \citenamefont {Fedorov}, \citenamefont {Chang}, \citenamefont {Jeng}, \citenamefont {Chuang},\ and\ \citenamefont {Chiang}}]{liu2023dirac}%
  \BibitemOpen
  \bibfield  {author} {\bibinfo {author} {\bibfnamefont {R.-Y.}\ \bibnamefont {Liu}}, \bibinfo {author} {\bibfnamefont {A.}~\bibnamefont {Huang}}, \bibinfo {author} {\bibfnamefont {R.}~\bibnamefont {Sankar}}, \bibinfo {author} {\bibfnamefont {J.~A.}\ \bibnamefont {Hlevyack}}, \bibinfo {author} {\bibfnamefont {C.-C.}\ \bibnamefont {Su}}, \bibinfo {author} {\bibfnamefont {S.-C.}\ \bibnamefont {Weng}}, \bibinfo {author} {\bibfnamefont {M.-K.}\ \bibnamefont {Lin}}, \bibinfo {author} {\bibfnamefont {P.}~\bibnamefont {Chen}}, \bibinfo {author} {\bibfnamefont {C.-M.}\ \bibnamefont {Cheng}}, \bibinfo {author} {\bibfnamefont {J.~D.}\ \bibnamefont {Denlinger}}, \bibinfo {author} {\bibfnamefont {S.-K.}\ \bibnamefont {Mo}}, \bibinfo {author} {\bibfnamefont {A.~V.}\ \bibnamefont {Fedorov}}, \bibinfo {author} {\bibfnamefont {C.-S.}\ \bibnamefont {Chang}}, \bibinfo {author} {\bibfnamefont {H.-T.}\ \bibnamefont {Jeng}}, \bibinfo {author} {\bibfnamefont {T.-M.}\ \bibnamefont {Chuang}},\ and\ \bibinfo {author} {\bibfnamefont
  {T.-C.}\ \bibnamefont {Chiang}},\ }\bibfield  {title} {\bibinfo {title} {Dirac nodal line in hourglass semimetal nb$_3$site$_6$},\ }\href {https://doi.org/10.1021/acs.nanolett.2c03293} {\bibfield  {journal} {\bibinfo  {journal} {Nano Letters}\ }\textbf {\bibinfo {volume} {23}},\ \bibinfo {pages} {380} (\bibinfo {year} {2023})}\BibitemShut {NoStop}%
\bibitem [{\citenamefont {Xie}\ \emph {et~al.}(2021{\natexlab{a}})\citenamefont {Xie}, \citenamefont {Gao}, \citenamefont {Xu}, \citenamefont {Zhang}, \citenamefont {Hu}, \citenamefont {Gao},\ and\ \citenamefont {Law}}]{Xie2021}%
  \BibitemOpen
  \bibfield  {author} {\bibinfo {author} {\bibfnamefont {Y.~M.}\ \bibnamefont {Xie}}, \bibinfo {author} {\bibfnamefont {X.~J.}\ \bibnamefont {Gao}}, \bibinfo {author} {\bibfnamefont {X.~Y.}\ \bibnamefont {Xu}}, \bibinfo {author} {\bibfnamefont {C.~P.}\ \bibnamefont {Zhang}}, \bibinfo {author} {\bibfnamefont {J.~X.}\ \bibnamefont {Hu}}, \bibinfo {author} {\bibfnamefont {J.~Z.}\ \bibnamefont {Gao}},\ and\ \bibinfo {author} {\bibfnamefont {K.~T.}\ \bibnamefont {Law}},\ }\bibfield  {title} {\bibinfo {title} {{Kramers nodal line metals}},\ }\href {https://doi.org/10.1038/s41467-021-22903-9} {\bibfield  {journal} {\bibinfo  {journal} {Nature Communications}\ }\textbf {\bibinfo {volume} {12}},\ \bibinfo {pages} {1} (\bibinfo {year} {2021}{\natexlab{a}})}\BibitemShut {NoStop}%
\bibitem [{\citenamefont {Wieder}\ \emph {et~al.}(2018)\citenamefont {Wieder}, \citenamefont {Bradlyn}, \citenamefont {Wang}, \citenamefont {Cano}, \citenamefont {Kim}, \citenamefont {Kim}, \citenamefont {Rappe}, \citenamefont {Kane},\ and\ \citenamefont {Bernevig}}]{wieder2018wallpaper}%
  \BibitemOpen
  \bibfield  {author} {\bibinfo {author} {\bibfnamefont {B.~J.}\ \bibnamefont {Wieder}}, \bibinfo {author} {\bibfnamefont {B.}~\bibnamefont {Bradlyn}}, \bibinfo {author} {\bibfnamefont {Z.}~\bibnamefont {Wang}}, \bibinfo {author} {\bibfnamefont {J.}~\bibnamefont {Cano}}, \bibinfo {author} {\bibfnamefont {Y.}~\bibnamefont {Kim}}, \bibinfo {author} {\bibfnamefont {H.-S.~D.}\ \bibnamefont {Kim}}, \bibinfo {author} {\bibfnamefont {A.~M.}\ \bibnamefont {Rappe}}, \bibinfo {author} {\bibfnamefont {C.~L.}\ \bibnamefont {Kane}},\ and\ \bibinfo {author} {\bibfnamefont {B.~A.}\ \bibnamefont {Bernevig}},\ }\bibfield  {title} {\bibinfo {title} {Wallpaper fermions and the nonsymmorphic dirac insulator},\ }\href {https://doi.org/10.1126/science.aan2802} {\bibfield  {journal} {\bibinfo  {journal} {Science}\ }\textbf {\bibinfo {volume} {361}},\ \bibinfo {pages} {246} (\bibinfo {year} {2018})}\BibitemShut {NoStop}%
\bibitem [{\citenamefont {Zhang}\ \emph {et~al.}(2019)\citenamefont {Zhang}, \citenamefont {Jiang}, \citenamefont {Song}, \citenamefont {Huang}, \citenamefont {He}, \citenamefont {Fang}, \citenamefont {Weng},\ and\ \citenamefont {Fang}}]{zhang2019catalogue}%
  \BibitemOpen
  \bibfield  {author} {\bibinfo {author} {\bibfnamefont {T.}~\bibnamefont {Zhang}}, \bibinfo {author} {\bibfnamefont {Y.}~\bibnamefont {Jiang}}, \bibinfo {author} {\bibfnamefont {Z.}~\bibnamefont {Song}}, \bibinfo {author} {\bibfnamefont {H.}~\bibnamefont {Huang}}, \bibinfo {author} {\bibfnamefont {Y.}~\bibnamefont {He}}, \bibinfo {author} {\bibfnamefont {Z.}~\bibnamefont {Fang}}, \bibinfo {author} {\bibfnamefont {H.}~\bibnamefont {Weng}},\ and\ \bibinfo {author} {\bibfnamefont {C.}~\bibnamefont {Fang}},\ }\bibfield  {title} {\bibinfo {title} {Catalogue of topological electronic materials},\ }\href@noop {} {\bibfield  {journal} {\bibinfo  {journal} {Nature}\ }\textbf {\bibinfo {volume} {566}},\ \bibinfo {pages} {475} (\bibinfo {year} {2019})}\BibitemShut {NoStop}%
\bibitem [{\citenamefont {Zhang}\ \emph {et~al.}(2020)\citenamefont {Zhang}, \citenamefont {Wu},\ and\ \citenamefont {Das~Sarma}}]{Zhang2020}%
  \BibitemOpen
  \bibfield  {author} {\bibinfo {author} {\bibfnamefont {R.}~\bibnamefont {Zhang}}, \bibinfo {author} {\bibfnamefont {F.}~\bibnamefont {Wu}},\ and\ \bibinfo {author} {\bibfnamefont {S.}~\bibnamefont {Das~Sarma}},\ }\bibfield  {title} {\bibinfo {title} {Möbius insulator and higher-order topology in $\mathbb{Z}_2$ topological antiferromagnets},\ }\href {https://doi.org/10.1103/PhysRevLett.124.136407} {\bibfield  {journal} {\bibinfo  {journal} {Physical Review Letters}\ }\textbf {\bibinfo {volume} {124}},\ \bibinfo {pages} {136407} (\bibinfo {year} {2020})}\BibitemShut {NoStop}%
\bibitem [{\citenamefont {Ishizaka}\ \emph {et~al.}(2011)\citenamefont {Ishizaka}, \citenamefont {Bahramy}, \citenamefont {Murakawa}, \citenamefont {Sakano}, \citenamefont {Shimojima}, \citenamefont {Sonobe}, \citenamefont {Koizumi}, \citenamefont {Shin}, \citenamefont {Miyahara}, \citenamefont {Kimura}, \citenamefont {Miyamoto}, \citenamefont {Okuda}, \citenamefont {Namatame}, \citenamefont {Taniguchi}, \citenamefont {Arita}, \citenamefont {Nagaosa}, \citenamefont {Kobayashi}, \citenamefont {Murakami}, \citenamefont {Kumai}, \citenamefont {Kaneko}, \citenamefont {Onose},\ and\ \citenamefont {Tokura}}]{ishizaka2011giant}%
  \BibitemOpen
  \bibfield  {author} {\bibinfo {author} {\bibfnamefont {K.}~\bibnamefont {Ishizaka}}, \bibinfo {author} {\bibfnamefont {M.~S.}\ \bibnamefont {Bahramy}}, \bibinfo {author} {\bibfnamefont {H.}~\bibnamefont {Murakawa}}, \bibinfo {author} {\bibfnamefont {M.}~\bibnamefont {Sakano}}, \bibinfo {author} {\bibfnamefont {T.}~\bibnamefont {Shimojima}}, \bibinfo {author} {\bibfnamefont {T.}~\bibnamefont {Sonobe}}, \bibinfo {author} {\bibfnamefont {K.}~\bibnamefont {Koizumi}}, \bibinfo {author} {\bibfnamefont {S.}~\bibnamefont {Shin}}, \bibinfo {author} {\bibfnamefont {H.}~\bibnamefont {Miyahara}}, \bibinfo {author} {\bibfnamefont {A.}~\bibnamefont {Kimura}}, \bibinfo {author} {\bibfnamefont {K.}~\bibnamefont {Miyamoto}}, \bibinfo {author} {\bibfnamefont {T.}~\bibnamefont {Okuda}}, \bibinfo {author} {\bibfnamefont {H.}~\bibnamefont {Namatame}}, \bibinfo {author} {\bibfnamefont {M.}~\bibnamefont {Taniguchi}}, \bibinfo {author} {\bibfnamefont {R.}~\bibnamefont {Arita}}, \bibinfo {author} {\bibfnamefont {N.}~\bibnamefont
  {Nagaosa}}, \bibinfo {author} {\bibfnamefont {K.}~\bibnamefont {Kobayashi}}, \bibinfo {author} {\bibfnamefont {Y.}~\bibnamefont {Murakami}}, \bibinfo {author} {\bibfnamefont {R.}~\bibnamefont {Kumai}}, \bibinfo {author} {\bibfnamefont {Y.}~\bibnamefont {Kaneko}}, \bibinfo {author} {\bibfnamefont {Y.}~\bibnamefont {Onose}},\ and\ \bibinfo {author} {\bibfnamefont {Y.}~\bibnamefont {Tokura}},\ }\bibfield  {title} {\bibinfo {title} {Giant rashba-type spin splitting in bulk bitei},\ }\href {https://doi.org/10.1038/nmat3051} {\bibfield  {journal} {\bibinfo  {journal} {Nature Materials}\ }\textbf {\bibinfo {volume} {10}},\ \bibinfo {pages} {521} (\bibinfo {year} {2011})}\BibitemShut {NoStop}%
\bibitem [{\citenamefont {Manchon}\ \emph {et~al.}(2015)\citenamefont {Manchon}, \citenamefont {Koo}, \citenamefont {Nitta}, \citenamefont {Frolov},\ and\ \citenamefont {Duine}}]{manchon2015new}%
  \BibitemOpen
  \bibfield  {author} {\bibinfo {author} {\bibfnamefont {A.}~\bibnamefont {Manchon}}, \bibinfo {author} {\bibfnamefont {H.~C.}\ \bibnamefont {Koo}}, \bibinfo {author} {\bibfnamefont {J.}~\bibnamefont {Nitta}}, \bibinfo {author} {\bibfnamefont {S.~M.}\ \bibnamefont {Frolov}},\ and\ \bibinfo {author} {\bibfnamefont {R.~A.}\ \bibnamefont {Duine}},\ }\bibfield  {title} {\bibinfo {title} {New perspectives for rashba spin–orbit coupling},\ }\href {https://doi.org/10.1038/nmat4360} {\bibfield  {journal} {\bibinfo  {journal} {Nature Materials}\ }\textbf {\bibinfo {volume} {14}},\ \bibinfo {pages} {871} (\bibinfo {year} {2015})}\BibitemShut {NoStop}%
\bibitem [{\citenamefont {Bradley}\ and\ \citenamefont {Cracknell}(1972)}]{bradley1972symmetry}%
  \BibitemOpen
  \bibfield  {author} {\bibinfo {author} {\bibfnamefont {C.~J.}\ \bibnamefont {Bradley}}\ and\ \bibinfo {author} {\bibfnamefont {A.~P.}\ \bibnamefont {Cracknell}},\ }\href@noop {} {\emph {\bibinfo {title} {The Mathematical Theory of Symmetry in Solids: Representation Theory for Point Groups and Space Groups}}}\ (\bibinfo  {publisher} {Clarendon Press},\ \bibinfo {address} {Oxford},\ \bibinfo {year} {1972})\BibitemShut {NoStop}%
\bibitem [{\citenamefont {Dresselhaus}\ \emph {et~al.}(2008)\citenamefont {Dresselhaus}, \citenamefont {Dresselhaus},\ and\ \citenamefont {Jorio}}]{dresselhaus2008group}%
  \BibitemOpen
  \bibfield  {author} {\bibinfo {author} {\bibfnamefont {M.~S.}\ \bibnamefont {Dresselhaus}}, \bibinfo {author} {\bibfnamefont {G.}~\bibnamefont {Dresselhaus}},\ and\ \bibinfo {author} {\bibfnamefont {M.}~\bibnamefont {Jorio}},\ }\href {https://doi.org/10.1007/978-3-540-32899-5} {\emph {\bibinfo {title} {Group Theory: Application to the Physics of Condensed Matter}}},\ \bibinfo {edition} {1st}\ ed.\ (\bibinfo  {publisher} {Springer Berlin Heidelberg},\ \bibinfo {address} {Berlin, Heidelberg},\ \bibinfo {year} {2008})\ pp.\ \bibinfo {pages} {XIV, 582},\ \bibinfo {note} {springer-Verlag Berlin Heidelberg, eBook ISBN: 978-3-540-32899-5, Published: 18 December 2007}\BibitemShut {NoStop}%
\bibitem [{\citenamefont {Zhao}\ \emph {et~al.}(2020)\citenamefont {Zhao}, \citenamefont {Nakamura}, \citenamefont {Arras}, \citenamefont {Paillard}, \citenamefont {Chen}, \citenamefont {Gosteau}, \citenamefont {Li}, \citenamefont {Yang},\ and\ \citenamefont {Bellaiche}}]{PhysRevLett.125.216405}%
  \BibitemOpen
  \bibfield  {author} {\bibinfo {author} {\bibfnamefont {H.~J.}\ \bibnamefont {Zhao}}, \bibinfo {author} {\bibfnamefont {H.}~\bibnamefont {Nakamura}}, \bibinfo {author} {\bibfnamefont {R.}~\bibnamefont {Arras}}, \bibinfo {author} {\bibfnamefont {C.}~\bibnamefont {Paillard}}, \bibinfo {author} {\bibfnamefont {P.}~\bibnamefont {Chen}}, \bibinfo {author} {\bibfnamefont {J.}~\bibnamefont {Gosteau}}, \bibinfo {author} {\bibfnamefont {X.}~\bibnamefont {Li}}, \bibinfo {author} {\bibfnamefont {Y.}~\bibnamefont {Yang}},\ and\ \bibinfo {author} {\bibfnamefont {L.}~\bibnamefont {Bellaiche}},\ }\bibfield  {title} {\bibinfo {title} {Purely cubic spin splittings with persistent spin textures},\ }\href {https://doi.org/10.1103/PhysRevLett.125.216405} {\bibfield  {journal} {\bibinfo  {journal} {Phys. Rev. Lett.}\ }\textbf {\bibinfo {volume} {125}},\ \bibinfo {pages} {216405} (\bibinfo {year} {2020})}\BibitemShut {NoStop}%
\bibitem [{\citenamefont {Basak}\ \emph {et~al.}(2011)\citenamefont {Basak}, \citenamefont {Lin}, \citenamefont {Wray}, \citenamefont {Xu}, \citenamefont {Fu}, \citenamefont {Hasan},\ and\ \citenamefont {Bansil}}]{PhysRevB.84.121401}%
  \BibitemOpen
  \bibfield  {author} {\bibinfo {author} {\bibfnamefont {S.}~\bibnamefont {Basak}}, \bibinfo {author} {\bibfnamefont {H.}~\bibnamefont {Lin}}, \bibinfo {author} {\bibfnamefont {L.~A.}\ \bibnamefont {Wray}}, \bibinfo {author} {\bibfnamefont {S.-Y.}\ \bibnamefont {Xu}}, \bibinfo {author} {\bibfnamefont {L.}~\bibnamefont {Fu}}, \bibinfo {author} {\bibfnamefont {M.~Z.}\ \bibnamefont {Hasan}},\ and\ \bibinfo {author} {\bibfnamefont {A.}~\bibnamefont {Bansil}},\ }\bibfield  {title} {\bibinfo {title} {Spin texture on the warped dirac-cone surface states in topological insulators},\ }\href {https://doi.org/10.1103/PhysRevB.84.121401} {\bibfield  {journal} {\bibinfo  {journal} {Phys. Rev. B}\ }\textbf {\bibinfo {volume} {84}},\ \bibinfo {pages} {121401} (\bibinfo {year} {2011})}\BibitemShut {NoStop}%
\bibitem [{\citenamefont {Bychkov}\ and\ \citenamefont {Rashba}(1984)}]{Bychkov1984}%
  \BibitemOpen
  \bibfield  {author} {\bibinfo {author} {\bibfnamefont {Y.~A.}\ \bibnamefont {Bychkov}}\ and\ \bibinfo {author} {\bibfnamefont {E.~I.}\ \bibnamefont {Rashba}},\ }\bibfield  {title} {\bibinfo {title} {{Oscillatory effects and the magnetic susceptibility of carriers in inversion layers}},\ }\href {https://doi.org/10.1088/0022-3719/17/33/015} {\bibfield  {journal} {\bibinfo  {journal} {Journal of Physics C: Solid State Physics}\ }\textbf {\bibinfo {volume} {17}},\ \bibinfo {pages} {6039} (\bibinfo {year} {1984})}\BibitemShut {NoStop}%
\bibitem [{\citenamefont {100}\ \emph {et~al.}(1954)\citenamefont {100}, \citenamefont {Er},\ and\ \citenamefont {Dresselhaust}}]{1001954}%
  \BibitemOpen
  \bibfield  {author} {\bibinfo {author} {\bibfnamefont {V.~E.}\ \bibnamefont {100}}, \bibinfo {author} {\bibfnamefont {N.~B.}\ \bibnamefont {Er}},\ and\ \bibinfo {author} {\bibfnamefont {G.}~\bibnamefont {Dresselhaust}},\ }\bibfield  {title} {\bibinfo {title} {{Spin-Orbit Coupling Effects in Zinc Blende Structures}},\ }\href@noop {} {\bibfield  {journal} {\bibinfo  {journal} {Phys. Rev. Phys. Rev. Phys. Rev. Z. Naturforsch. Z. Naturforsch. Phys. Rev}\ }\textbf {\bibinfo {volume} {96}},\ \bibinfo {pages} {280} (\bibinfo {year} {1954})}\BibitemShut {NoStop}%
\bibitem [{\citenamefont {Dyakonov}(2017)}]{Dyakonov2017}%
  \BibitemOpen
  \bibinfo {editor} {\bibfnamefont {M.~I.}\ \bibnamefont {Dyakonov}},\ ed.,\ \href {https://doi.org/10.1007/978-3-319-65436-2} {\emph {\bibinfo {title} {Spin Physics in Semiconductors}}},\ \bibinfo {series} {Springer Series in Solid-State Sciences}, Vol.\ \bibinfo {volume} {157}\ (\bibinfo  {publisher} {Springer},\ \bibinfo {address} {Cham, Switzerland},\ \bibinfo {year} {2017})\BibitemShut {NoStop}%
\bibitem [{\citenamefont {Xie}\ \emph {et~al.}(2021{\natexlab{b}})\citenamefont {Xie}, \citenamefont {Gao}, \citenamefont {Xu}, \citenamefont {Zhang}, \citenamefont {Hu}, \citenamefont {Gao},\ and\ \citenamefont {Law}}]{xie2021kramers}%
  \BibitemOpen
  \bibfield  {author} {\bibinfo {author} {\bibfnamefont {Y.-M.}\ \bibnamefont {Xie}}, \bibinfo {author} {\bibfnamefont {X.-J.}\ \bibnamefont {Gao}}, \bibinfo {author} {\bibfnamefont {X.~Y.}\ \bibnamefont {Xu}}, \bibinfo {author} {\bibfnamefont {C.-P.}\ \bibnamefont {Zhang}}, \bibinfo {author} {\bibfnamefont {J.-X.}\ \bibnamefont {Hu}}, \bibinfo {author} {\bibfnamefont {J.~Z.}\ \bibnamefont {Gao}},\ and\ \bibinfo {author} {\bibfnamefont {K.~T.}\ \bibnamefont {Law}},\ }\bibfield  {title} {\bibinfo {title} {Kramers nodal line metals},\ }\href {https://doi.org/10.1038/s41467-021-22903-9} {\bibfield  {journal} {\bibinfo  {journal} {Nature Communications}\ }\textbf {\bibinfo {volume} {12}},\ \bibinfo {pages} {3064} (\bibinfo {year} {2021}{\natexlab{b}})}\BibitemShut {NoStop}%
\bibitem [{\citenamefont {Belik}\ \emph {et~al.}(2006)\citenamefont {Belik}, \citenamefont {Stefanovich}, \citenamefont {Lazoryak},\ and\ \citenamefont {Takayama-Muromachi}}]{Belik2006}%
  \BibitemOpen
  \bibfield  {author} {\bibinfo {author} {\bibfnamefont {A.~A.}\ \bibnamefont {Belik}}, \bibinfo {author} {\bibfnamefont {S.~Y.}\ \bibnamefont {Stefanovich}}, \bibinfo {author} {\bibfnamefont {B.~I.}\ \bibnamefont {Lazoryak}},\ and\ \bibinfo {author} {\bibfnamefont {E.}~\bibnamefont {Takayama-Muromachi}},\ }\bibfield  {title} {\bibinfo {title} {{BiInO3: A polar oxide with GdFeO3-type perovskite structure}},\ }\href {https://doi.org/10.1021/cm052627s} {\bibfield  {journal} {\bibinfo  {journal} {Chemistry of Materials}\ }\textbf {\bibinfo {volume} {18}},\ \bibinfo {pages} {1964} (\bibinfo {year} {2006})}\BibitemShut {NoStop}%
\bibitem [{\citenamefont {Bl{\"{o}}chl}(1994)}]{paw}%
  \BibitemOpen
  \bibfield  {author} {\bibinfo {author} {\bibfnamefont {P.~E.}\ \bibnamefont {Bl{\"{o}}chl}},\ }\bibfield  {title} {\bibinfo {title} {{Projector augmented-wave method}},\ }\href {https://doi.org/10.1103/PhysRevB.50.17953} {\bibfield  {journal} {\bibinfo  {journal} {Phys. Rev. B}\ }\textbf {\bibinfo {volume} {50}},\ \bibinfo {pages} {17953} (\bibinfo {year} {1994})}\BibitemShut {NoStop}%
\bibitem [{\citenamefont {Kresse}\ and\ \citenamefont {Furthm{\"{u}}ller}(1996)}]{vasp1}%
  \BibitemOpen
  \bibfield  {author} {\bibinfo {author} {\bibfnamefont {G.}~\bibnamefont {Kresse}}\ and\ \bibinfo {author} {\bibfnamefont {J.}~\bibnamefont {Furthm{\"{u}}ller}},\ }\bibfield  {title} {\bibinfo {title} {{Efficient iterative schemes for ab initio total-energy calculations using a plane-wave basis set}},\ }\href {https://doi.org/10.1103/PhysRevB.54.11169} {\bibfield  {journal} {\bibinfo  {journal} {Physical Review B}\ }\textbf {\bibinfo {volume} {54}},\ \bibinfo {pages} {11169} (\bibinfo {year} {1996})}\BibitemShut {NoStop}%
\bibitem [{\citenamefont {Kresse}\ and\ \citenamefont {Joubert}(1999)}]{vasp2}%
  \BibitemOpen
  \bibfield  {author} {\bibinfo {author} {\bibfnamefont {G.}~\bibnamefont {Kresse}}\ and\ \bibinfo {author} {\bibfnamefont {D.}~\bibnamefont {Joubert}},\ }\bibfield  {title} {\bibinfo {title} {{From ultrasoft pseudopotentials to the projector augmented-wave method}},\ }\href {https://doi.org/10.1103/PhysRevB.59.1758} {\bibfield  {journal} {\bibinfo  {journal} {Physical Review B}\ }\textbf {\bibinfo {volume} {59}},\ \bibinfo {pages} {1758} (\bibinfo {year} {1999})}\BibitemShut {NoStop}%
\bibitem [{\citenamefont {Perdew}\ \emph {et~al.}(1996)\citenamefont {Perdew}, \citenamefont {Burke},\ and\ \citenamefont {Ernzerhof}}]{pbe}%
  \BibitemOpen
  \bibfield  {author} {\bibinfo {author} {\bibfnamefont {J.~P.}\ \bibnamefont {Perdew}}, \bibinfo {author} {\bibfnamefont {K.}~\bibnamefont {Burke}},\ and\ \bibinfo {author} {\bibfnamefont {M.}~\bibnamefont {Ernzerhof}},\ }\bibfield  {title} {\bibinfo {title} {{Generalized Gradient Approximation Made Simple}},\ }\href {https://doi.org/10.1103/PhysRevLett.77.3865} {\bibfield  {journal} {\bibinfo  {journal} {Phys. Rev. Lett.}\ }\textbf {\bibinfo {volume} {77}},\ \bibinfo {pages} {3865} (\bibinfo {year} {1996})}\BibitemShut {NoStop}%
\bibitem [{\citenamefont {Bl{\"{o}}chl}\ \emph {et~al.}(1994)\citenamefont {Bl{\"{o}}chl}, \citenamefont {Jepsen},\ and\ \citenamefont {Andersen}}]{BlochlPRB94T}%
  \BibitemOpen
  \bibfield  {author} {\bibinfo {author} {\bibfnamefont {P.~E.}\ \bibnamefont {Bl{\"{o}}chl}}, \bibinfo {author} {\bibfnamefont {O.}~\bibnamefont {Jepsen}},\ and\ \bibinfo {author} {\bibfnamefont {O.~K.}\ \bibnamefont {Andersen}},\ }\bibfield  {title} {\bibinfo {title} {{Improved tetrahedron method for Brillouin-zone integrations}},\ }\href {https://doi.org/10.1103/PhysRevB.49.16223} {\bibfield  {journal} {\bibinfo  {journal} {Phys. Rev. B}\ }\textbf {\bibinfo {volume} {49}},\ \bibinfo {pages} {16223} (\bibinfo {year} {1994})}\BibitemShut {NoStop}%
\bibitem [{\citenamefont {Wu}\ \emph {et~al.}(2018)\citenamefont {Wu}, \citenamefont {Zhang}, \citenamefont {Song}, \citenamefont {Troyer},\ and\ \citenamefont {Soluyanov}}]{WU2017}%
  \BibitemOpen
  \bibfield  {author} {\bibinfo {author} {\bibfnamefont {Q.}~\bibnamefont {Wu}}, \bibinfo {author} {\bibfnamefont {S.}~\bibnamefont {Zhang}}, \bibinfo {author} {\bibfnamefont {H.-F.}\ \bibnamefont {Song}}, \bibinfo {author} {\bibfnamefont {M.}~\bibnamefont {Troyer}},\ and\ \bibinfo {author} {\bibfnamefont {A.~A.}\ \bibnamefont {Soluyanov}},\ }\bibfield  {title} {\bibinfo {title} {Wanniertools : An open-source software package for novel topological materials},\ }\href {https://doi.org/https://doi.org/10.1016/j.cpc.2017.09.033} {\bibfield  {journal} {\bibinfo  {journal} {Computer Physics Communications}\ }\textbf {\bibinfo {volume} {224}},\ \bibinfo {pages} {405 } (\bibinfo {year} {2018})}\BibitemShut {NoStop}%
\bibitem [{\citenamefont {Mostofi}\ \emph {et~al.}(2014)\citenamefont {Mostofi}, \citenamefont {Yates}, \citenamefont {Pizzi}, \citenamefont {Lee}, \citenamefont {Souza}, \citenamefont {Vanderbilt},\ and\ \citenamefont {Marzari}}]{MOSTOFI20142309}%
  \BibitemOpen
  \bibfield  {author} {\bibinfo {author} {\bibfnamefont {A.~A.}\ \bibnamefont {Mostofi}}, \bibinfo {author} {\bibfnamefont {J.~R.}\ \bibnamefont {Yates}}, \bibinfo {author} {\bibfnamefont {G.}~\bibnamefont {Pizzi}}, \bibinfo {author} {\bibfnamefont {Y.-S.}\ \bibnamefont {Lee}}, \bibinfo {author} {\bibfnamefont {I.}~\bibnamefont {Souza}}, \bibinfo {author} {\bibfnamefont {D.}~\bibnamefont {Vanderbilt}},\ and\ \bibinfo {author} {\bibfnamefont {N.}~\bibnamefont {Marzari}},\ }\bibfield  {title} {\bibinfo {title} {An updated version of wannier90: A tool for obtaining maximally-localised wannier functions},\ }\href {https://doi.org/https://doi.org/10.1016/j.cpc.2014.05.003} {\bibfield  {journal} {\bibinfo  {journal} {Computer Physics Communications}\ }\textbf {\bibinfo {volume} {185}},\ \bibinfo {pages} {2309} (\bibinfo {year} {2014})}\BibitemShut {NoStop}%
\bibitem [{\citenamefont {Aroyo}\ \emph {et~al.}(2006)\citenamefont {Aroyo}, \citenamefont {Perez-Mato}, \citenamefont {Capillas}, \citenamefont {Kroumova}, \citenamefont {Ivantchev}, \citenamefont {Madariaga}, \citenamefont {Kirov},\ and\ \citenamefont {Wondratschek}}]{aroyo2006bilbao}%
  \BibitemOpen
  \bibfield  {author} {\bibinfo {author} {\bibfnamefont {M.~I.}\ \bibnamefont {Aroyo}}, \bibinfo {author} {\bibfnamefont {J.~M.}\ \bibnamefont {Perez-Mato}}, \bibinfo {author} {\bibfnamefont {C.}~\bibnamefont {Capillas}}, \bibinfo {author} {\bibfnamefont {E.}~\bibnamefont {Kroumova}}, \bibinfo {author} {\bibfnamefont {S.}~\bibnamefont {Ivantchev}}, \bibinfo {author} {\bibfnamefont {G.}~\bibnamefont {Madariaga}}, \bibinfo {author} {\bibfnamefont {A.}~\bibnamefont {Kirov}},\ and\ \bibinfo {author} {\bibfnamefont {H.}~\bibnamefont {Wondratschek}},\ }\bibfield  {title} {\bibinfo {title} {Bilbao crystallographic server: I. databases and crystallographic computing programs},\ }\href@noop {} {\bibfield  {journal} {\bibinfo  {journal} {Zeitschrift f{\"u}r Kristallographie-Crystalline Materials}\ }\textbf {\bibinfo {volume} {221}},\ \bibinfo {pages} {15} (\bibinfo {year} {2006})}\BibitemShut {NoStop}%
\bibitem [{\citenamefont {Soluyanov}\ and\ \citenamefont {Vanderbilt}(2011)}]{soluyanov2011computing}%
  \BibitemOpen
  \bibfield  {author} {\bibinfo {author} {\bibfnamefont {A.~A.}\ \bibnamefont {Soluyanov}}\ and\ \bibinfo {author} {\bibfnamefont {D.}~\bibnamefont {Vanderbilt}},\ }\bibfield  {title} {\bibinfo {title} {Computing topological invariants without inversion symmetry},\ }\href@noop {} {\bibfield  {journal} {\bibinfo  {journal} {Physical Review B—Condensed Matter and Materials Physics}\ }\textbf {\bibinfo {volume} {83}},\ \bibinfo {pages} {235401} (\bibinfo {year} {2011})}\BibitemShut {NoStop}%
\bibitem [{\citenamefont {Nye}(1985)}]{nye1985physical}%
  \BibitemOpen
  \bibfield  {author} {\bibinfo {author} {\bibfnamefont {J.~F.}\ \bibnamefont {Nye}},\ }\href@noop {} {\emph {\bibinfo {title} {Physical properties of crystals: their representation by tensors and matrices}}}\ (\bibinfo  {publisher} {Oxford university press},\ \bibinfo {year} {1985})\BibitemShut {NoStop}%
\bibitem [{\citenamefont {Sirotin}\ and\ \citenamefont {Shaskol}(1982)}]{sirotin1982fundamentals}%
  \BibitemOpen
  \bibfield  {author} {\bibinfo {author} {\bibfnamefont {I.}~\bibnamefont {Sirotin}}\ and\ \bibinfo {author} {\bibfnamefont {M.~P.}\ \bibnamefont {Shaskol}},\ }\href@noop {} {\emph {\bibinfo {title} {Fundamentals of crystal physics}}}\ (\bibinfo  {publisher} {Mir Publishers},\ \bibinfo {year} {1982})\BibitemShut {NoStop}%
\bibitem [{\citenamefont {Bandyopadhyay}\ and\ \citenamefont {Dasgupta}(2021)}]{PhysRevB.103.014105}%
  \BibitemOpen
  \bibfield  {author} {\bibinfo {author} {\bibfnamefont {S.}~\bibnamefont {Bandyopadhyay}}\ and\ \bibinfo {author} {\bibfnamefont {I.}~\bibnamefont {Dasgupta}},\ }\bibfield  {title} {\bibinfo {title} {Orbital-dependent spin textures in ferroelectric rashba systems},\ }\href {https://doi.org/10.1103/PhysRevB.103.014105} {\bibfield  {journal} {\bibinfo  {journal} {Phys. Rev. B}\ }\textbf {\bibinfo {volume} {103}},\ \bibinfo {pages} {014105} (\bibinfo {year} {2021})}\BibitemShut {NoStop}%
\end{thebibliography}%
\end{document}